\documentclass[12pt]{article}

\def\ZZ {{\bf Z}}
\def\RR {{\bf R}}

\def\Ztwo{{\ZZ}_{2}}
\def\Zthree{{\ZZ}_{3}}

\def\tr{{\rm tr}}
\def\Tr{{\rm Tr}}

\def\etal{{\it et al.}}
\def\ie{{\it i.e.}}

\usepackage{fullpage,epsfig,fancyheadings}

\newcommand{\beq}{\begin{equation}}
\newcommand{\eeq}{\end{equation}}

\begin{document}
\vspace*{-.6in}
\thispagestyle{empty}

\begin{flushright}
CALT-68-2065\\
hep-th/9607201
\end{flushright}
\baselineskip = 20pt

\vspace{.5in}

{\Large
\begin{center}
Lectures on Superstring and M Theory Dualities\footnote{Work 
supported in part by
the U.S. Dept. of Energy under Grant No. DE-FG03-92-ER40701.}
\end{center}}

\vspace{.4in}

\begin{center}
John H. Schwarz\\
\emph{California Institute of Technology, Pasadena, CA  91125, USA}
\end{center}
\vspace{1in}

\begin{center}

\textbf{Abstract}

\end{center}

\begin{quotation}
\noindent These lectures begin by reviewing the evidence for S duality of the
toroidally compactified heterotic string in 4d that was obtained in the period 1992--94.
Next they review recently discovered dualities that relate all five
of the 10d superstring theories and a quantum extension of 11d supergravity called
M theory. The study of $p$-branes of various dimensions (some of which are $D$-branes)
plays a central role. The final sections survey supersymmetric string vacua in 6d and
some of the dual constructions by which they can be obtained. 
Special emphasis is given to a class of  $N=1$ models that exhibit ``heterotic-heterotic duality.''
\end{quotation}
\vfil
\centerline{\it Lectures presented at the ICTP  Spring School (March 1996)}
\centerline{\it and at the TASI  Summer School (June 1996)}
\bigskip
\newpage
\pagenumbering{arabic}

\section{Introduction}

In the first superstring revolution (1984--85) we learned that there are just
five superstring theories, each of which admits a 10d
Poincar\'e-invariant vacuum and has a perturbation expansion that is consistent
at every finite order.\cite{green87} Three of the theories have $N=1$ supersymmetry in 10d
(type I and the two heterotic theories), 
one has $N=2$ supersymmetry in 10d with the two supercharges having
opposite chirality (type IIA) and one has $N=2$ supersymmetry in 10d with the
two supersymmetries having the same chirality (type IIB).  
One of the theories is based on unoriented open and closed strings (type I)
and the other four are based on oriented closed strings. In short, each of the five
theories appears to be quite different from the others, with very distinctive features.
Of course, we don't really want five theories, since there is only one universe to
explain. So the hope  that I and others expressed in the mid 1980's
was that some of these might turn out to be equivalent or inconsistent, but it
wasn't apparent how this could happen.

In the second superstring revolution (1994--??) we are learning that all of these
different superstring theories are consistent, but that they are
non-perturbatively equivalent. Each of them represents a perturbation expansion of a
single underlying theory about a distinct point in the moduli space of quantum vacua.
Moreover, there is a sixth rather special point in this moduli space characterized by
an 11d Poincar\'e-invariant vacuum. The rules for doing quantum
mechanics in the 11d vacuum are not yet understood, but  the answer (whatever it is)
has been named `M theory'. Some people believe that M theory is more
fundamental than the five superstring theories in 10d, but I do not share that
viewpoint. Rather, for reasons that will be explained in these lectures,
I believe that it is on a roughly equal footing with the type IIB superstring theory (or
`F theory'). Each of these descriptions (extended by various possible compactifications)
gives access to different `patches'
of the space of quantum vacua. A good analogy, which I heard first from Vafa 
(at the CERN workshop in June 1996), is that the underlying
theory is being defined in much the same way that one defines a manifold. A manifold
can be defined by giving a covering by open sets  that are 
diffeomorphic to open sets in $\RR^n$
and by consistently defining transition functions on their overlaps.
In the proposed analogy, each of the superstring
theories corresponds to one of the open sets, and the dualities that characterize their
non-perturbative equivalences correspond to the transition functions. From this viewpoint
the various dualities could be viewed as part of the {\it definition} of the underlying theory
rather than as conjectured theorems that require proof. Of course, as in the
case of ordinary transition functions, they must satisfy a number of consistency
conditions. Exactly what consistency conditions are required has not yet been carefully
spelled out. There is a widespread belief that a deeper formulation of the theory ought to
exist, and I tend to share that belief, but it is also conceivable that the various
perturbation expansions and non-perturbative dualities constitute the
best definition of the theory.

\subsection{Duality Symmetries of Supergravity and Superstring Theories}

There are three types of dualities that appear in superstring theory, which go by the names
of $S$, $T$, and $U$. Two theories, call them A and B, are said to be $S$ dual if theory A
at strong coupling is equivalent to theory B at weak coupling (and vice versa). This means
that there is an exact map between the A and B descriptions that includes, among other
things the relation $\phi_A = - \phi_B$. Here $\phi_A$ and $\phi_B$ denote the
respective dilaton fields, which determine the string coupling constant $\lambda$ according to
the rule $\lambda = {\rm exp} <\phi>$. Theories A and B are called $T$ dual if theory A
compactified on a space of large volume is equivalent to theory B
compactified on a space of small volume (and vice versa).  This means,
for example, that some other scalar
field $t$, the exponential of whose vev determines the volume of the compactified dimensions,
satisfies $t_A =  - t_B$. $T$ dualities
can be checked order-by-order in string perturbation theory, and therefore they
were the first ones to be understood. Theories A and B can be called $U$ dual
if theory A compactified on a space of large (or small) volume is equivalent to theory B
at strong (or weak) coupling.  In this case $t_A = \pm \phi_B$.
This is not exactly the definition of $U$ duality that was originally
proposed, but I feel it is in the spirit of the original proposal and  find it
to be convenient. When present, each
of these dualities is supposed to constitute an exact quantum equivalence,
which means that the two `theories' should really be viewed as different
descriptions of a single theory. It sometimes happens that a single theory is self-dual
under a group of these dualities. In this case, the dualities are symmetries --- discrete gauge
symmetries, to be precise. This means that configurations related by duality
transformations describe equivalent vacua, which should be identified as one and the same.

The appearance of a non-compact global symmetry group $G$ is
a characteristic feature of the supergravity theories that
represent the low-energy effective action for the massless modes of a superstring
compactification. Typically, the group $G$ is realized nonlinearly by scalar fields that
parametrize the homogeneous space $G/H$, where $H$ is the maximal compact
subgroup of $G$.  The first example of this phenomenon, with $G = SL(2,\RR)$ and $H =
U(1)$, was uncovered in 1976 in a version of $N = 4$  4d supergravity by Cremmer,
Ferrara,  and Scherk.\cite{cremmer78a}  Curiously,
a discrete subgroup of the symmetry of this particular example corresponds precisely to
the example of $S$ duality that was first recognized in string theory -- that of the toroidally
compactified heterotic string.  An analogous non-compact $E_7$ symmetry was
found in $N = 8$ 4d supergravity by Cremmer and Julia in
1978,\cite{cremmer78b}  and many other examples were worked out
thereafter.\cite{salam89} The Cremmer--Julia example corresponds to the
toroidally compactified type II string, and combines $S$, $T$, and $U$
dualities in a single discrete group. (As mentioned above, this usage of the term
`$U$ duality' differs a bit from the one proposed by Hull and Townsend,\cite{hull94}
which refers to the entire  group as `$U$ duality.')

The first proposal for the non-perturbative behavior of string theory
was the 1990 suggestion of Font \etal\cite{font90} that
the $SL(2,\ZZ)$ subgroup of the $SL(2,\RR)$ of Cremmer, Ferrara, and Scherk
should be an exact symmetry of the
heterotic string toroidally compactified (in the way described by Narain\cite{narain86}) to
4d.  They named this discrete symmetry group $S$ duality, because
the $N=1$ superfield (containing the axion and dilaton) that parametrizes
$SL(2,\RR)/U(1)$ is often called $S$.  That $S$ duality should be an exact
symmetry of the quantum string theory
was a bold conjecture, since a $\Ztwo$ subgroup is an
electric-magnetic duality in which the coupling constant is inverted
($g_{el} \rightarrow g_{mag} \propto
1/g_{el}$).  Thus, it relates the strong coupling limit to a weakly coupled description. 
This proposal extends the duality conjecture of Montonen and Olive~\cite{montonen77}
from supersymmetric gauge theories to the superstring setting.

Since string theory had only been formulated in perturbation theory, 
the proposal of Font {\it et al.}, when it first appeared, seemed to me
to be an intriguing but untestable suggestion.  
In any case, the $S$ duality conjecture was eventually picked up
and pursued by Sen and myself.\cite{sen92,schwarz92,schwarz93}
As we will see, non-trivial tests of $S$ duality have been formulated and verified.
The technical tool that makes it possible to extract non-perturbative information
about theories that have only been defined perturbatively is supersymmetry.
Specifically, when there is enough supersymmetry,  states
belonging  to `short representations' of the supersymmetry algebra are
exactly stable and have many of their properties protected from quantum
corrections -- both perturbative and non-perturbative. This will be discussed
in more detail later. 

$T$ duality, unlike $S$ duality, holds order by order in
string perturbation theory.\cite{giveon94}
In the simplest case -- compactification on a circle -- the
group is $\Ztwo$ and the transformation corresponds to inversion of the radius
$(R \rightarrow \alpha'/R)$.  Of course, as mentioned above,
$R$ is determined by the value of a scalar field (a $T$ modulus). As in the case of $S$ duality,
when $T$ duality is a symmetry of a single theory, it is a discrete gauge symmetry 
that is realized as a field
transformation, whereas when it relates two apparently different theories
it is a field identification. 

\subsection{The 4d Heterotic String}

In the example of toroidal compactification of the
heterotic string no supersymmetry is broken, and in 4d there are 132 scalar
fields that live on the Narain moduli space ${\cal M}_{6,22}$. Narain spaces
${\cal M}_{k,l}$ are defined by
\begin{equation}
{\cal M}_{k,l} = SO(k,l;\ZZ) \backslash SO(k,l) / SO(k) \times SO(l).
\end{equation}
It is convenient to introduce this notation here, since we shall encounter various Narain spaces
in the course of these lectures. Recall that $SO(k,l)$ is the noncompact form
of $SO(k+l)$ that preserves a metric with $k$ plus signs and $l$ minus signs. The group
$SO(k)\times SO(l)$ is its maximal compact subgroup and the quotient space
$SO(k,l) /SO (k) \times  SO(l)$ is a homogeneous space of dimension $kl$.
The discrete group $SO(k,l;\ZZ)$ is an infinite group consisting of all $SO(k,l)$
matrices with integer entries. When $l = k + 16$, it is the subgroup of $SO(k,l)$ that preserves a
certain even self-dual lattice of signature ($k, l$) introduced by Narain. 
A homogeneous space is very smooth and
well-behaved, but modding out by the discrete group introduces orbifold singularities, corresponding to the fixed points of the discrete group, in the moduli space.
The $T$ duality group for the 4d heterotic string is $G_T = SO(6, 22; \ZZ)$, and
the 132 scalar fields belong to 22 Abelian $N = 4$ gauge multiplets. In terms
of compactification from 10d, 21 of the scalars originate from the metric,
15 from the two-form $B_{\mu\nu}$, and 96 from the 16 $U(1)$ gauge fields that
form the Cartan subalgebra of $E_8 \times E_8$ or $SO(32)$.

The toroidally compactified heterotic string also contains two additional
scalar fields -- called the axion $\chi$ and the dilaton $\phi$ -- which belong
to the $N = 4$ supergravity multiplet.  The dilaton is the 10d dilaton shifted
by a function of the other moduli such that the exponential of
its vev gives the 4d coupling constant. The 4d axion is the scalar field that
is dual to the  two-form $B$ in 4d.
The supergravity theory that contains these fields is precisely
the one studied by Cremmer, Ferrara, and Scherk. They showed that
$\chi$ and $\phi$  parametrize the homogeneous space 
$SL(2, \RR)/U(1)$.  Actually, in the quantum theory, only the
discrete $S$ duality subgroup $SL(2,\ZZ)$ is a symmetry, and the
moduli space is
\begin{equation}
{\cal M}_S  = SL(2,\ZZ)\backslash SL(2, \RR)/U(1) .
\end{equation}
To see how this works, let us introduce a complex scalar field
\begin{equation}
\rho = \chi + i e^{-2\phi} = \rho_1 + i \rho_2 ~
\end{equation}
whose vev is $<\rho > = {\theta/ 2\pi} + { i/
\lambda^2}$, where $\theta$ is the vacuum angle and $\lambda$ is the coupling
constant. $N=4$ Yang--Mills theories have vanishing
$\beta$ function, so that $\theta$ and $\lambda$ are well-defined independent
of scale.  In terms of $\rho$, the $SL(2,\ZZ)$ symmetry is realized by the
non-linear transformations
\begin{equation}
\rho \rightarrow {a \rho + b\over c\rho + d} ~, \quad \quad
\left(\begin{array} {cc} a & b  \\ c & d \end{array}\right)\, \in \, SL(2,  \ZZ) .
\end{equation}
As usual, when instanton effects are taken into account,
the continuous Peccei--Quinn symmetry $\chi \rightarrow \chi + b$,
is broken to the discrete subgroup for which $b$ is an integer.  This subgroup
and the inversion $\rho \rightarrow - 1/\rho$ generate the discrete group
$SL(2,\ZZ)$ or, when matrices are not distinguished from their negatives, $PSL(2,\ZZ)$.
When $\theta =0$, a special case is inversion of the coupling constant
$\lambda \rightarrow 1/\lambda$. In general, the $SL(2,\ZZ)$ symmetry of the
theory is broken completely by any specific choice of vacuum.  Only when the vev
of $\rho$ is at one of the orbifold points of the moduli space does some
unbroken symmetry ($\Ztwo$ or $\Zthree$) remain.

Mathematically, $S$ and $T$ duality are quite analogous in the 4d low-energy
effective field theory, even though their implications for string theory
are dramatically different.  This analogy was one of the original motivations
for proposing that $S$ duality should also be a symmetry.
The massless bosonic fields of the toroidally compactified heterotic string are
the metric tensor $g_{\mu\nu}$, the axion-dilaton field $\rho$,
28 Abelian gauge fields $A_\mu^a$ (6 from the 10d metric, 6 from the 10d
two-form, and 16 from the Cartan subalgebra), and the 132 moduli
parametrizing ${\cal M}_{6,22}$.  These are the only massless bosonic fields
at generic points in the classical moduli space.  At the singular points,
where there is enhanced gauge symmetry, there are more.  The 132 moduli
are conveniently
described as a symmetric $28 \times 28$ matrix belonging to the group
$SO(6,22)$:
\begin{equation}
M^T = M, ~~ M^T LM = L \end{equation}
\begin{equation}
L = \left(\begin{array} {ccc} 0 & I_6 & 0 \cr I_6 & 0 & 0 \cr 0 & 0 & I_{22}
\end{array}\right).
\end{equation}
Under a $T$ duality transformation given by an $SO(6,22;\ZZ)$ matrix $\Omega$
satisfying $\Omega^T L \Omega = L$
\begin{equation}
 M \rightarrow \Omega M \Omega^T, \quad A_\mu \rightarrow \Omega A_\mu {}~,
\end{equation}
while $g_{\mu\nu}$ and $\rho$ are invariant.

The 28 U(1) gauge fields $A_\mu^a$ give rise to 28  electric and 28
magnetic charges.  A convenient way to define them is to assume that space-time
is asymptotically flat and use the asymptotic behavior of the
field strengths:
\begin{equation}
 F_{0i}^a \sim {q_{el}^a\over r^3} x^i~~\quad
\tilde{F}_{0i}^a \sim {q_{mag}^a\over r^3} x^i ~.\end{equation}
The allowed charges are controlled by the
asymptotic values of the moduli ($\rho \sim \rho^{(0)}$ and $M_{ab} \sim
M_{ab}^{(0)}$) and a pair of vectors $\alpha^a, \beta^a$ belonging to the
Narain lattice, which is an even self-dual Lorentzian lattice of signature
(6,22).  The formulas are
\begin{equation}
 q_{el}^a = {1\over \rho_2^{(0)}} M_{ab}^{(0)} (\alpha^b + \rho_1^{(0)}
\beta^b), \quad q^a_{mag} = L_{ab}\beta^b ~.\end{equation}
These formulas automatically incorporate the
Dirac--Schwinger--Zwanziger--Witten quantization rules ({\it i.e.}, the
quantization condition for dyons in the presence of a $\theta$ angle). States
in the perturbative string spectrum carry electric charge only and therefore
have $\beta^a =0$.

\subsection{The 4d Type II Superstring}

The type II (A or B) superstring compactified on $T^6$ is approximated at
low-energy by $N=8$ supergravity. The classical theory has a non-compact
symmetry
group $E_{7,7}$. The natural conjecture for the duality group in this case is
the discrete subgroup $E_7(\ZZ)$, which is defined as the intersection of
the continuous $E_{7,7}$ group and the discrete group $Sp(28;\ZZ)$.\cite{hull94}
Written in the 56-dimensional
fundamental representation, it is evident that $E_{7,7}$ is a subgroup of
the non-compact group $Sp(28)$. (Later, in other contexts, the symbol $Sp(n)$
will represent a compact group.) The analog of the Narain moduli space in this case is
\begin{equation}
{\cal M}  = E_7(\ZZ)\backslash E_{7,7}/SU(8).
\end{equation}
The scalar fields of $N=8$ supergravity parametrize this 70-dimensional space.
As in the $N=4$ heterotic theory, there are once again 28 $U(1)$ gauge fields.
However, this time only 12 of their electric charges are excited (by Kaluza--Klein and
winding excitations) in the perturbative string spectrum. The remaining 16 electric charges
and all of the magnetic charges are only carried by non-perturbative excitations.
One way of understanding this is to note the decomposition
\begin{equation}
E_7(\ZZ) \supset SO(6,6;\ZZ) \times SL(2,\ZZ),
\end{equation}
which exhibits the $T$ duality and $S$ duality subgroups.  With respect to this subgroup,
the fundamental {\bf 56} representation decomposes as ${\bf 56} = ({\bf 12}, {\bf 2})
+ ({\bf 32}, {\bf 1} )$. The 12 electric charges that occur perturbatively are
carried by states whose mass is finite at weak coupling in the string metric.
They are associated to the first term, as are the dual magnetic charges, which give
states whose mass is proportional to $1/\lambda^2$. The 16 electric and magnetic charges
associated to the 32-dimensional spinorial representation of $SO(6,6)$ turn out to be
carried by $D$-branes, and as a result they give masses proportional to $1/\lambda$.

\subsection{The BPS Condition}

The $N$-extended 4d supersymmetry algebra (in 2-component notation)
includes the anticommutator
\begin{equation}
\{ Q^I_{\alpha}, Q^J_{\beta} \} = \epsilon_{\alpha\beta} Z^{I J}.
\end{equation}
The $N(N-1)/2$ central charges $Z^{IJ} = - Z^{JI}$
are complex numbers whose real and imaginary parts give the electric and
magnetic charges
associated with the $N(N-1)/2$ $U(1)$ gauge fields in the $N$-extended 4d
supergravity multiplet. The supersymmetry algebra
implies that the mass
of any state is bounded below by its central charges.  This bound, known as the
Bogomol'nyi bound, is very important. When the mass of a state attains the minimum
value allowed for given charges (and moduli), the state is said to be BPS saturated.
BPS states belong to smaller representations of the algebra than are possible when
the bound is not saturated. There are actually several possibilities for how this can be
achieved.  To explain this, it is convenient to make an $SO(N)$ change of basis
such that (in the case of $N=4$, for example)
\begin{equation}
Z = \left(\begin{array}{cccc}0 & Z_1 & 0 & 0\cr -Z_1& 0 & 0 & 0 \cr 0& 0 & 0 & Z_2 \cr
0 & 0 & -Z_2 & 0 \end{array}\right) ~.
\end{equation}
Thus we see that in the $N=4$ case, even though the supergravity multiplet has
six $U(1)$ gauge fields, a generic configuration can be described by only considering
two electric and two magnetic charges. In this case there are two ways to achieve
BPS saturation. In the first case,
the mass satisfies the relations $M = |Z_1| = |Z_2|$. This gives `ultrashort'
multiplets, such as the 16-dimensional gauge multiplet. The second possibility
for a BPS state is $M = |Z_1| > |Z_2|$. The first case occurs when the electric charge
vector $\alpha^a$ and the magnetic charge vector $\beta^a$ are parallel, while
in the second case they are not parallel. Since BPS states in the perturbative string
spectrum are purely electric, they are necessarily of the first type.

These considerations are important in making comparisons of string states and
black holes. Static extremal black hole configurations with $M = |Z_1| = |Z_2|$
turn out to preserve one-half of the supersymmetry and to have a horizon of vanishing
area (and hence no Bekenstein--Hawking entropy). Ones with $M = |Z_1| > |Z_2|$,
on the other hand, preserve only one-quarter of the supersymmetry and have a
horizon of finite area. There are analogous statements that can be made in
the $N=8$ case. In that case, in order to obtain a finite-area horizon, it is necessary that
$M$ equals only one of the four $|Z_i|$'s so that seven-eighths of the supersymmetry is broken.
There has been dramatic progress recently in accounting for the entropy
of supersymmetric black holes with finite area horizons in
terms of the counting of microscopic string degrees of freedom. However, I will
leave that (and generalizations) to other lecturers.

\subsection{Tests of S Duality}

$T$ duality works
perturbatively and is well understood, but how can we prove $S$ duality without
knowing non-perturbative string theory?  As yet, we cannot prove it, but we can
subject the conjecture to some non-trivial tests by focusing on BPS
states.  The essential fact, pointed out long ago by Witten and Olive,\cite{witten78}
is that such states can receive no quantum corrections -- perturbative or
nonperturbative -- to their masses so long as the supersymmetry remains
unbroken.  Thus, a non-trivial prediction of $S$ duality,
which we can attempt to check, is that the multiplicities of BPS states
are $SL(2, \ZZ)$ invariant.  Note that since the vacuum breaks $S$ and $T$
duality spontaneously, the BPS states do
not form degenerate multiplets.

Let us now explore which states in the elementary string spectrum 
of the 4d heterotic string saturate the
Bogomol'nyi bound. Absorbing the moduli $M^{(0)}$ in the definition of the
Narain lattice, the BPS condition for purely electric states becomes
\begin{equation}
({\rm Mass})^2 = {1\over 16 \rho_2^{(0)}} \hat{\alpha}^a
(I + L)_{ab} \hat{\alpha}^b =
{1\over 8\rho_2^{(0)}} (\hat{\alpha}_R)^2 ~,\end{equation}
where $\hat{\alpha} L \hat{\alpha} = \hat{\alpha}_R \cdot \hat{\alpha}_R -
\hat{\alpha}_L \cdot \hat{\alpha}_L$.  ($\hat{\alpha}_L$ is 22-dimensional and
$\hat{\alpha}_R$ is 6-dimensional.  They correspond to the left-moving and
right-moving internal momenta of the string.)  Now we should compare the  free
string spectrum, which is given by
\begin{equation}
({\rm Mass})^2 = {1\over 4\rho_2^{(0)}} \left[ {1\over 2}
(\hat{\alpha}_L)^2 + N_L - 1\right]
= {1\over 4\rho_2^{(0)}} \left[ {1\over 2} (\hat{\alpha}_R)^2 + N_R -
\delta\right] ~.
\end{equation}
$N_L$ and $N_R$ represent left-moving and right-moving oscillator excitations.
The parameter $\delta$ is 1/2 in the NS sector and $0$ in the R sector.
Alternatively, it is simply 0 in the GS
formulation. The factor of $(\rho_2^{(0)})^{-1}$ appears because the mass is
computed with respect to the canonically normalized Einstein  metric.  It does
not appear if one uses the string metric, which differs by a dilaton-dependent
Weyl rescaling.  The Einstein metric is more natural in the present context,
because it is invariant under $S$ duality transformations.  Comparing formulas,
one sees that the Bogomol'nyi bound is saturated provided that $N_R =
\delta$ (which gives eight bosonic and eight fermionic right-moving modes -- the short
representation of $N = 4$) and $N_L = 1 + {1\over 2} \hat{\alpha} L
\hat{\alpha}$.  Thus, if $\hat{\alpha} L \hat{\alpha} = 2n - 2$, for a
non-negative integer $n$, then there is a short $N = 4$ multiplet for every
solution of $N_L = n$.  These states are only ``electrically'' charged.  The
challenge is to find their predicted $S$ duality partners.  Specifically, every
elementary string excitation of the type we have just described $(\vec{\alpha}
= \vec{\ell}$, $\vec{\beta} = 0)$ should have magnetically charged partners with
$\vec{\alpha} = a \vec{\ell}$ and $\vec{\beta} = c \vec{\ell}$. Since $a$ and
$c$ are elements of an $SL(2,\ZZ)$ matrix, they are relatively prime
integers.

Sen has investigated the partners of electrically charged states with
$\hat{\alpha} L \hat{\alpha} = - 2 $ ({\it i.e.}, $N_L = 0$).\cite{sen94a}  He has shown that
$S$ duality partners with $c = 1$
can be identified with BPS monopole solutions
(and their dyonic generalizations) of the effective field theory.
These solutions saturate the bound, of
course.  Thus, as we have explained, they should persist with exactly this mass
in the complete quantum
string theory.  For $c > 1$, Sen argued that one should examine
multi-BPS dyon bound states.  Specifically, he showed that the prediction of $S$ duality is
that each multi-BPS dyon moduli space should admit a unique normalizable harmonic
form.  Poincar\'e duality would give a second one unless it is self-dual or
anti-self-dual.  He constructed such an anti-self-dual
form explicitly for the case of $c =2$,\cite{sen94b}
providing the first really non-trivial test of $S$ duality.
Progress toward extending this result to $c > 2$ has been made by Segal and Selby \cite{segal}
and by Porrati.\cite{porrati95} More recently, a simpler and more general proof has
been constructed \cite{landsteiner} using D-brane techniques. \cite{bershadsky}.

\section{DUALITIES IN NINE DIMENSIONS}

\subsection{Introductory Comments}

The five 10d superstring theories  -- types I, IIA, IIB and the $E_8 \times
E_8$ and $SO(32)$ heterotic -- are related to one another by a rich variety of
dualities.  The dualities that require compactification of only one spatial dimension, leaving a
9d Minkowski space-time, are sufficient to show that  all five are
related to one another.  This strongly suggests that they are best regarded
as different descriptions of a single underlying theory.  Each one is better suited
to describing some portion of the moduli space of possible vacua than the others.  
In a later section, we will discuss some of the additional dualities that emerge upon
compactification to 6d, but in this section we wish to explore what can be
learned while retaining an uncompactified $\RR^{\, 9}$.

Two of the relevant dualities are $T$ dualities, which can be understood
perturbatively, and therefore they were understood prior to the recent
non-perturbative discoveries.  When the IIA and IIB theories are each
compactified on a circle, so that altogether the space-time topology is $\RR^{ 9}
\times S^1$, the two theories are $T$ dual.\cite{dai89,dine89}  This means that they describe
identical physics, provided that the radius of one circle is the inverse of the
other one (in string units).  In a similar manner, one can show that the $SO(32)$
and $E_8 \times E_8$ heterotic string theories are $T$ dual when each of them
is compactified on a circle.\cite{narain86,ginsparg87}  
This heterotic case is somewhat more subtle than the
type II one.  Wilson lines have to be included, as part of the
characterization of the compactification, in order to match
corresponding points in the moduli space of
9d vacua.  The relevant moduli space in this case is the Narain
space ${\cal M}_{1,17}$ defined earlier.

The pair of $T$ dualities described above provides two connections among the
five superstring theories.  To see that all five are connected requires
examining non-perturbative dualities -- analogs of the $S$ duality of the 4d
heterotic string  discussed in Section 1.
One way of addressing the problem is to ask, for each of the five theories, whether the
strong coupling limit has a dual weakly coupled description.  As in the case of
the $S$ duality of the 4d heterotic string, the procedure is to identify a
plausible candidate for an $S$ dual description and then to examine its
consequences.  The result is that an interesting and consistent story emerges.
In fact, it is so compelling that there can be little doubt about the truth of the proposed
dualities.  So let me now say what they are.

The type I and $SO(32)$ heterotic string theories are $S$ dual.\cite{witten95a,polchinski96}  
This means that the respective dilatons are related by $\phi_I = - \phi_H$, so that the
coupling constants $(\lambda = e^{<\phi>})$ are reciprocal to one another.
This identification is supported by the fact that both have the same low-energy
effective field theory descriptions ($N = 1$ supergravity coupled to $SO(32)$
super Yang--Mills in 10d). The field
redefinition $\phi_I = - \phi_H$ must be accompanied by a Weyl rescaling of the metric
to convert from one version of the action to the other.  On the other hand, the type IIB
superstring in 10d is self-dual.\cite{hull94}  More precisely, there is an $SL(2,\ZZ)$
$S$-duality group, very much like that of the 4d heterotic string, that is
a  gauge symmetry of the theory.  This will be described in detail later.

The strong coupling limits of the IIA and $E_8 \times E_8$ theories turn out to
provide quite a different surprise.  In each case there is an eleventh
dimension (tenth spatial dimension) that becomes large at strong 
coupling.\cite{townsend95a,witten95a,horava95}
Specifically, the size of this dimension scales as $L_{11} \sim
\lambda^{2/3}$, where $\lambda$ is the 10d coupling constant.  Such
a compact dimension is completely invisible in perturbation theory (an expansion about
$\lambda = 0$), which is why it passed unnoticed for so many years.  In the
IIA case, the hidden dimension is a circle $S^1$, whereas in the $E_8 \times
E_8$ case it is a line interval $I$, or (more precisely) an $S^1/\ZZ_2$ orbifold.
This means that in the $E_8 \times E_8$ case one can visualize the space-time
as an 11d space-time with two 10d faces, which are
sometimes referred to as ``end-of-the-world 9-branes,'' since they have nine
spatial dimensions.\cite{horava95}  One of the $E_8$ gauge groups is associated to each face.
In any case, at strong coupling the faces move apart and (away from the faces)
the theory is described by the same 11d bulk theory that describes
the IIA theory at strong coupling.  This 11d theory is described in
leading order in a low-energy expansion by 11d supergravity, a
classical field theory that was discovered almost 20 years ago.\cite{cremmer78c}  It is not yet
known what is the correct algorithm that determines all the higher-dimension
terms of the low-energy expansion of the effective action, 
but since we are confident that there is 
a consistent quantum theory, such an algorithm should exist.  The unknown
11d quantum theory is referred to as M  theory.  As we will
discuss, certain of its supersymmetric solitons are known, 
and they provide a handle on many of its interesting properties.

The equivalences discussed above can be summarized by the diagram in Figure 1.
This diagram is sufficient to show that the five superstring theories are all
part of a single structure, but it is by no means the whole story.  There are a
variety of other surprising dualities that are only revealed upon
compactification of additional dimensions.  Some of these will be  described
later.
\begin{figure}[t]
\centerline{\epsfxsize=4truein \epsfbox{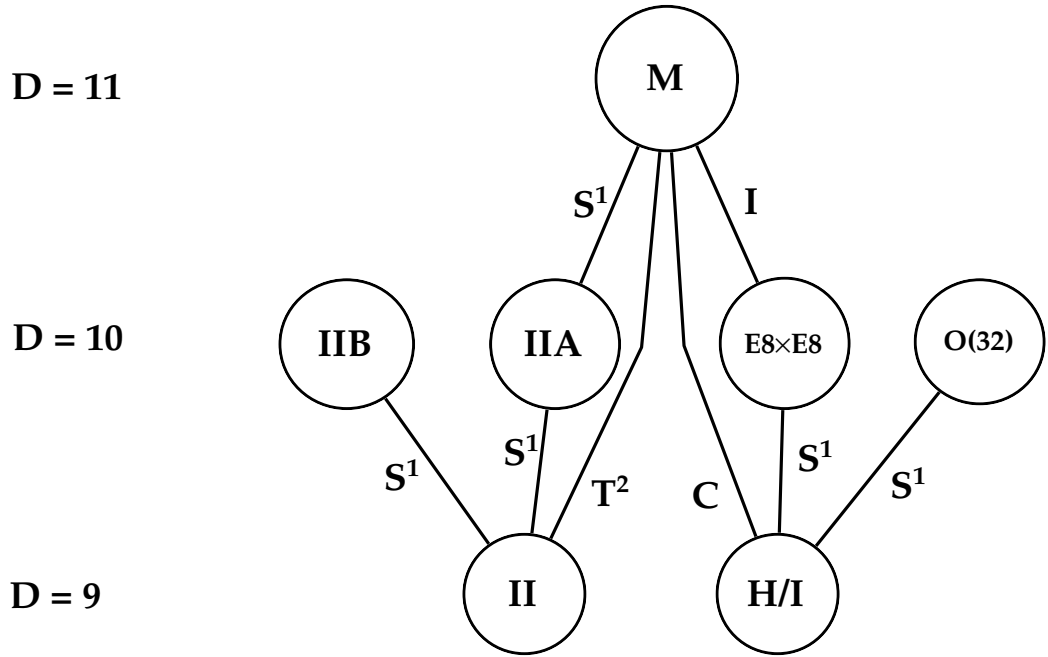} }
\caption{Duality Connections.}
\end{figure}

The IIA/IIB $T$ duality and the IIA/M
$S$ duality can be combined as a duality between IIB theory on $\RR^{9} \times
S^1$ and M  theory on $\RR^{ 9} \times T^2$.  This viewpoint turns out to be very
powerful for understanding the structure of both the IIB and M  theories
separately, as will be discussed in considerable detail.  The $T^2$ is
characterized by three real parameters -- its area $A_M$ and its modular
parameter $\tau$, which characterizes its complex structure up to an
$SL(2,\ZZ)$ transformation.  Indeed, we will find that in 9d the
$SL(2,\ZZ)$ modular group of the torus precisely corresponds to the $SL(2,\ZZ)$
$S$ duality group of the IIB theory.  This geometrization of $S$ duality is quite profound.

There is an analogous duality relating M  theory and the $SO(32)$ theory (both
type I and heterotic).  This duality can be understood as arising as a corollary of
the first one after modding out by a suitable $\ZZ_2$ symmetry.
In this case, M  theory compactified on a cylinder $C =
I \times S^1$ is dual to the $SO(32)$ theory compactified on a circle $S^1$.
This also has consequences for both theories.

\subsection{General Features of $p$-branes}

M  theory and the various superstring theories admit a rich variety of soliton
solutions.  When the core of such a solution extends over  $p$ spatial dimensions
(and one time dimension),
the soliton is called a $p$-brane.  Of special interest are $p$-brane solitons
that saturate a BPS bound, which means that they preserve some fraction of the underlying
supersymmetry.  We will focus on ones that preserve one-half of the
supersymmetry, but ones that preserve a smaller fraction, such as one-quarter
or one-eighth, can be constructed. 

Supersymmetric $p$-branes are a natural generalization of the BPS states
($0$-branes) discussed in Section 1.  In this case the relevant central
charges in the supersymmetry algebra are $p$-forms.\cite{townsend95b} 
Their magnitudes give a
lower bound on the $p$-brane tension $T_p$, which is the mass per unit volume of the brane.
Supersymmetry is preserved when the bound is saturated.  As in the case of the
$0$-branes, the BPS condition ensures that solutions of classical low-energy
supergravity field equations exhibit some features of the exact 
quantum string theory, such as the relationship between the tension and
the charge.

The supergravity solutions are non-singular in certain cases, so that the energy is smoothly 
spread over a region surrounding a $p$-dimensional subspace. In other
cases there are delta function singularities at the core that can be compensated
by postulating the presence of ``fundamental $p$-branes.''   Many authors
distinguish these two categories of $p$-branes by calling them ``solitonic''
and ``fundamental,'' respectively.  As far as I can tell, this is a distinction
that need not persist in the underlying quantum theory, rather it could just be
an artifact of formalism and approximations.  Therefore, we regard both
categories of $p$-brane solutions as ``solitonic'' without focussing on this
distinction. (See Ref.~\cite{duff94} for a review of $p$-brane solutions.)

The effective supergravities in question contain various antisymmetric tensor
gauge fields.  These can be represented as differential forms
\begin{equation}
A_n = A_{\mu_{1}\mu_{2} \ldots \mu_{n}} dx^{\mu_{1}}~\wedge dx^{\mu_{2}}
\wedge \ldots \wedge dx^{\mu_{n}} .
\end{equation}
In this notation, a gauge transformation is given by $\delta A_n = d \Lambda_{n
- 1}$, and the gauge-invariant field strength is $F_{n + 1} = dA_n$.  When
interactions are included, these formulas are sometimes modified.  The origin
of $p$-branes can be understood by considering an action that (schematically)
has the structure~\cite{horowitz91}
\begin{equation}
S \sim \int d^D x \sqrt{-g} \{R + (\partial \phi)^2 + e^{-a\phi} F_{n+1}^2
+ \ldots\}.
\end{equation}
Here, $\phi$ represents a dilaton field, $R$ is the scalar curvature, and $a$
is a numerical constant whose value depends on the particular theory.  The dots
include all the additional terms required to make the theory locally
supersymmetric.  In this case it is meaningful to seek BPS
$p$-brane solutions, and it turns out that solutions exist for $p = n - 1$ and
$p = D - n - 3$.  By a straightforward generalization of the nomenclature of
Maxwell theory, it is natural to call these ``electric'' and ``magnetic,''
respectively.  The electric $p$-brane, with $p = n - 1$, has an $n$-dimensional
world-volume.  The fact that it is a source for ``electric'' charge is
exhibited by the coupling
\begin{equation}
\int A_{\mu_{1} \ldots \mu_{n}} {\partial x^{\mu_{1}}\over \partial \sigma_1}
\ldots {\partial x^{\mu_{n}}\over\partial\sigma_n} d^n \sigma,
\end{equation}
which generalizes the familiar $j \cdot A$ coupling of Maxwell theory.

A $p$-brane in $D$ dimensions (let's assume it is an infinite hyperplane, for
simplicity) can be encircled by a $(D-p-2)$-dimensional sphere $S^{D-p-2}$.
Thus, the ``electric charge'' of the $p$-brane is given by a straightforward
generalization of Gauss's law for point charges
\begin{equation}
Q_E \sim \int_{S^{D - p - 2}} * F,
\end{equation}
where $*F$ is the Hodge dual of $F$.  In these lectures, we will not need to
commit ourselves to specific normalization conventions.  Similarly, a dual $(D
- p - 4)$-brane has ``magnetic charge''
\begin{equation}
Q_M \sim \int_{S^{p + 2}} F.
\end{equation}
Note that the charge associated with a $p$-brane has dimension $(length)^{D/2 -
2 - p}$.  
This is dimensionless when $p = (D-4)/2$ -- \ie, for point particles in
4d, strings in 6d, membranes in 8d, etc.  In these cases the
electric and magnetic branes have the same dimensionality and it is possible to
have dyonic $p$-branes.

The charges of $p$-branes can also be described by generalizations of Coulomb's
law.  So, for an electric $p$-brane, as $r\rightarrow\infty$
\begin{equation}
A \sim  {Q_E \over r^{D - p - 3}}\omega_{p + 1},
\end{equation}
where $r$ is the transverse distance from the brane and $\omega_{p+1}$ is the
volume form for the $p$-brane world-volume.  Similarly, for the dual magnetic
($D -p-4$)-brane, as $r\rightarrow\infty$
\begin{equation}
F \sim  {Q_M \over r^{p + 2}}\Omega_{p + 2},
\end{equation}
where $\Omega_{p+2}$ is the volume form on a sphere $S^{p + 2}$ surrounding the
brane.  In this case it is convenient to describe the magnetic field, rather
than the potential, in order to avoid introducing generalizations of Dirac
strings.  Of course, the distinction between electric and magnetic branes is
not so great, since it is often possible to make a duality transformation that
replaces $A$ by a dual potential $\tilde{A}$ whose field strength $d\tilde{A}$
is the dual of $F = dA$.  From the point of view of $\tilde{A}$, the original
electric brane is magnetic and vice versa.  Another significant fact,\cite{nepomechie85} 
noted more  than ten years ago, is that the Dirac quantization condition has a
straightforward generalization to the charges carried by a dual pair of
$p$-branes: $Q_E Q_M \in 2\pi \ZZ$.  This assumes appropriate normalization
conventions, of course.

The crudest first approximation to classical $p$-brane dynamics is given by a
straightforward generalization of the Nambu area formula
for the string world-sheet action.  This gives an action
proportional to the $(p + 1)$-dimensional volume induced by
embedding  the world volume into the $D$-dimensional target space:
\begin{equation}
S_{eff} = T_P \int \sqrt{{\rm det}\ G_{\alpha\beta}}\, d^{p + 1} \sigma ,
\end{equation}
where
\begin{equation}
G_{\alpha\beta} = \eta_{\mu\nu} \partial_\alpha x^\mu \partial_\beta
x^\nu,  \alpha,\beta = 0, 1, \ldots, p,
\end{equation}
and $\eta$ is the metric (Minkowski, for example) of the target space.  Just
as for strings, this formula is invariant under reparametrizations of the world
volume.  Also, it defines the $p$-brane tension $T_p$ -- the universal mass per
unit volume of the $p$-brane.  Note that $T_p \sim (mass)^{p +1}$.

\subsection{Specific $p$-branes}

Let us now examine what BPS $p$-branes occur in the theories of most interest to
us.  We begin with 11d supergravity, the low-energy effective field theory for
M  theory.  11d supergravity has three massless fields: the metric
$g_{\mu\nu}$ (with 44 physical polarizations), the gravitino $\psi_\mu$ (with
128 physical polarizations), and a three-form potential $C_{\mu\nu\rho}$ (with 84 physical
polarizations).  By the reasoning given above one expects to find two kinds of
branes associated with $C$:  an electric 2-brane~\cite{bergshoeff87,duff91} 
and a magnetic 5-brane,\cite{guven92} and
this is indeed the case, even though 11d supergravity has no
dilaton field. These branes have a number of interesting properties,
which we will return to later.

The only anomaly-free 10d theories with $N=1$ supersymmetry have as their massless
sector $N=1$ supergravity coupling to either $SO(32)$
or $E_8 \times E_8$ super Yang--Mills matter.\cite{green84}  
The relevant antisymmetric tensor gauge field
that couples to $p$-branes is the two-form potential $B$ belonging to the
supergravity multiplet.  In this case the electric $p$-brane is a $1$-brane, which
{\it is} the heterotic string.\cite{dabholkar95} 
Its magnetic dual is a $5$-brane.\cite{strominger90} Type I strings cannot
be found in this way, because they are not BPS (and hence not stable).

Type IIA supergravity in 10d can be understood as arising from compactification of M theory 
on a circle.\cite{townsend95a,witten95a} 
Doing this, the 11d metric gives rise to the 10d metric, a one-form $A$, and a
dilaton $\phi_A$.  Specifically, if $g^{(10)}$ denotes the IIA string metric,
\begin{equation}
g_{MN}^{(11)} dx^M dx^N = e^{-2\phi_{A}/3} g_{\mu\nu}^{(10)} dx^\mu dx^\nu
+ e^{4\phi_{A}/3} (dx^{11} - A_\mu dx^\mu)^2.
\end{equation}
Identifying the string coupling constant $\lambda = e^{<\phi_{A}>}$, one sees
that $R_{11} \sim \lambda^{2/3}$, as was asserted earlier.  Also, the 11d
three-form $C$ decomposes in 10d into a three-form $C$ and a two-form $B$.

The IIA theory has six kinds of $p$-branes:\footnote{Another possibility,
8-branes, will not be considered here. (See Ref.~\cite{bergshoeff96}.)}
$p = 0,6$ associated to $A$; $p =
1,5$ associated to $B$; $p = 2,4$ associated to $C$.  Some of these have a
simple interpretation in terms of the 2-brane and 5-brane in 11d.  The 2-brane
and 5-brane in 10d are given by a straight dimensional reduction and the
1-brane and 4-brane in 10d are given by a double dimensional reduction.  The
fact that a $p$-brane solution in $D$ dimensions implies that there is also one
in $D-1$ dimensions (for $p <D-3$) after compactification on a circle depends
crucially on the BPS property. This allows one to form an infinite periodic
array of parallel $p$-branes in $D$ dimensions, and then a periodic identification
gives a single $p$-brane in $D-1$ dimensions.  Double dimensional reduction is more
straightforward:  one dimension of the $p$-brane wraps around the
circular dimension of the space-time.  The $0$-brane and $6$-brane couple to the 
Kaluza--Klein gauge field $A$, and can therefore be called Kaluza--Klein
$p$-branes.  The charge carried by a KK $0$-brane is interpreted as momentum
in the 11th dimension.  Their role in 11d is simply to allow this
momentum to be excited.  The dual 6-brane, on the other hand, has a tension
that diverges in the decompactification limit.  Thus, there is no corresponding
soliton in 11d Minkowski space.

Let us now consider the most interesting case of all -- type IIB
superstrings in 10d.\cite{green82}  In the NS-NS sector there is a 2-form potential
$B^{(1)}$.  The fundamental IIB string is electrically charged with respect to
this field.  In addition, the R-R sector has a zero-form
$\chi$, a two-form $B^{(2)}$, and a four-form $A_4$.  The four-form $A_4$ has a
self-dual field strength $(dA_4 = * dA_4)$, something that is possible
only when the number of spatial dimensions minus the number of time dimensions
is a multiple of four.

Formally, a zero-form gives a $(-1)$-brane and a $7$-brane, both of which are
rather special.  A $(-1)$-brane has a point-like world volume.  After a Wick
rotation, it can be interpreted as a kind of instanton called a $D$-Instanton.  
Its magnetic dual, a
$7$-brane, is also special.  Whenever $p = D-3$, the presence of the brane gives rise to a
conical deficit angle in the geometry of the transverse plane, a fact that is
artfully exploited by F theory.\cite{vafa96a}  
Here, we will only consider branes with $p < D - 3$.
The four-form $A_4$ gives rise to a self-dual 3-brane.   It has an
identified electric and magnetic charge, because $\int_{S^{5}} F_5 =
\int_{S^{5}} * F_5$.  Thus, $A_4$ only gives one kind of $p$-brane.
Finally, we turn to the two forms $B_{\mu\nu}^{(1)}$ and
$B_{\mu\nu}^{(2)}$.  Each can couple to an electric $1$-brane or a
magnetic $5$-brane.  However, as we will argue, these 
1-branes or 5-branes can form bound states.
Thus, we will get an infinite family of strings labelled by two electric
charges $(q_1, q_2)$ and an infinite family dual magnetic $5$-branes labelled
by two magnetic charges.  

\subsection{Type IIB String Solitons}

As has already been noted, the type IIB superstring in 10d has two two-form
potentials, $B_{\mu\nu}^{(1)}$ and $B_{\mu\nu}^{(2)}$.  Therefore,
string-like solutions can, in general, carry a pair of charges
\begin{equation}
q_I \sim \int_{S^{7}} * dB^{(I)}, \quad I = 1,2.
\end{equation}
Let us construct these solutions explicitly.
To make the $SL(2,\RR)$ symmetry of the supergravity field equations manifest, it
is convenient to introduce a two-component vector notation
\begin{equation}
H = dB = \left(\begin{array}{cc}
dB^{(1)}\\
dB^{(2)} \end{array} \right).
\end{equation}
It is also convenient to combine the $RR$ scalar $\chi$ and the dilaton $\phi$
into a complex scalar field
\begin{equation}
\rho = \chi + i e^{-\phi},
\end{equation}
and to represent the vev of this field by
\begin{equation}
<\rho> = \rho_0= \chi_0 + i e^{-\phi_{0}} = {\theta\over 2\pi} +
{i\over\lambda_B}.
\end{equation}
The field $\rho$ is very similar to the axion-dilaton field of the 4d heterotic
theory described in Section 1.  Indeed, as in that case, it transforms 
nonlinearly under an $SL(2,\RR)$ transformation $\Lambda
= \left(\begin{array}{cc} a & b\\ c & d \end{array} \right)$, by the rule $\rho
\rightarrow {a\rho + b\over c\rho + d}$.  There are differences of detail,
however.  For example, the imaginary part of $\rho$ is $e^{-\phi}$ here, whereas
in Section 1 it was $e^{-2\phi}$.  It is also convenient to introduce the
symmetric $SL(2,\RR)$ matrix
\begin{equation}
{\cal M} = e^\phi \left(\begin{array}{cc}
|\rho|^2 & \chi\\
\chi & 1 \end{array} \right),
\end{equation}
which transforms by the simple rule
\begin{equation}
{\cal M} \rightarrow \Lambda {\cal M} \Lambda^T.
\end{equation}

The $B$ fields transform linearly by the rule $B \rightarrow (\Lambda^T)^{-1}
B$, while the canonical metric $g_{\mu\nu}$ and the four-form $A_4$ are
invariant.  Note that since the dilaton transforms, the IIB string metric
\begin{equation}
g_{\mu\nu}^{(B)} = e^{\phi/2} g_{\mu\nu},  \label{twog}
\end{equation}
is not $SL(2,\RR)$ invariant.  For this reason, it is convenient to use
the canonical metric for the time being.

The string-like solutions of the IIB supergravity field equations that we are
seeking have $A_4$ and all fermi fields equal to zero.  While it is difficult
to formulate a convenient action that gives the complete field equations
(because the field strength of $A_4$ is self-dual), it is not hard to find the action
that gives the field equations with $A_4$ and the fermi fields set equal to
zero.  It is
\begin{equation}
S = \int d^{10} x \sqrt{-g} (R - {1\over 12} H_{\mu\nu\rho}^T {\cal
M}H^{\mu\nu\rho} + {1\over 4} \tr (\partial^\mu {\cal M} \partial_\mu {\cal M}^{-1})).
\end{equation}
This action is manifestly invariant under global $SL(2,\RR)$ transformations.
The solution we seek consists of a string-like soliton along the $x^1$ axis with
$B$ charges $(q_1, q_2)$ and vacuum defined by $\rho (r) \sim  \rho_0$
as $ r \rightarrow\infty$, where $r^2 = \vec x \cdot \vec x$, and $\vec x$
refers to the eight transverse directions $x^2, x^3, \ldots, x^9$.  Also, the metric
should approach the Minkowski metric as $r \rightarrow \infty$.

Our problem was solved some time ago by Dabholkar, {\it et al.}, for a special case,
namely $\vec q = (1,0)$ and $\rho_0 = i$.\cite{dabholkar90} 
This solution, which has $\chi = 0$
and $B_{\mu\nu}^{(2)} = 0$, arose in considering the heterotic string, which
does not contain the fields $\chi$ and $B_{\mu\nu}^{(2)}$.  However, its
equations agree with the ones being considered here when they are set to zero.
Using the $SL(2,\RR)$ symmetry of the IIB theory, the solution of Dabholkar, {\it et
al.}, can be transformed to give a IIB solution with charges $(q_1,  q_2)$ 
and $\rho (r) {\sim } \rho_0$.\cite{schwarz95a}
The solution obtained in this way is given by
\[
ds^2 = A^{-3/4} (-dt^2 + (dx^1)^2) + A^{1/4} d\vec x \cdot d \vec x\]
\[
B_{01}^{(I)} = q_I \Delta_{(q_1,q_2)}^{-1/2} A^{-1}\]
\begin{equation}
\rho = {i (q_2 \chi_0 + q_1 |\rho_0|^2) A^{1/2} - q_2 e^{-\phi_{0}}\over
i(q_1 \chi_0 + q_2) A^{1/2} + q_1 e^{-\phi_{0}}},
\end{equation}
where
\begin{equation}
A = 1 + {Q \Delta_{(q_1,q_2)}^{1/2}\over r^6},
\end{equation}
\begin{equation}
\Delta_{(q_1,q_2)} = e^{\phi_{0}} |q_1 - q_2 \rho_0|^2, \label{twoa}
\end{equation}
and the charge $Q$ is a constant proportional to the tension scale $T_1^{(B)}$
and the 10d Newton constant.

While classically $q_1$ and $q_2$ are arbitrary real numbers, 
quantum mechanically they must be
integers.  This follows (by the same reasoning Dirac used to explain the
quantization of electric charge) from the existence of 5-branes and
the Dirac quantization condition.  Later, we will argue that stability requires
that $q_1$ and $q_2$ should actually be {\it relatively-prime integers}.  By allowing all
pairs of relatively prime integers, we define an infinite family of string-like
solitons, which form an irreducible $SL(2,\ZZ)$ multiplet.  Note that if, for a
given string solution, $\rho_0$ is analytically continued outside the fundamental region
${\cal F}$ of $SL(2,\ZZ)$, then the $SL(2,\ZZ)$ transformation that brings
$\rho_0$ back inside ${\cal F}$ will redefine the charges of the string.

By considering the asymptotic behavior of the metric component $g_{00}$ for $r
\rightarrow\infty$, one can read off the ``ADM tension'' of the string\cite{schwarz95a}
\begin{equation}
T_{(q_{1}, q_{2})} = \Delta_{(q_1,q_2)}^{1/2} T_1^{(B)}.   \label{twob}
\end{equation}
To get a sense of the meaning of this equation, it is convenient to restrict
to the special case $\chi_0 = 0$, so that $\rho_0 = i/\lambda_B$.  Then
the tension of the $(q_1, q_2)$ string in the canonical metric is
\begin{equation}
T_{(q_{1}, q_{2})} = (\lambda_B q_1^2 + \lambda_B^{-1} q_2^2)^{1/2} T_1^{(B)}.
\end{equation}
Converting to the IIB string metric, redefines this by a factor of
$\lambda_B^{-1/2}$, giving
\begin{equation}
\tilde{T}_{(q_{1},q_{2})} = (q_1^2 + \lambda_B^{-2} q_2^2)^{1/2} T_1^{(B)}.
\end{equation}
Thus, in the string metric, the fundamental string tension is a constant,
$\tilde{T}_{(1,0)} = T_1^{(B)}$.  The $D$-string, which carries $RR$ charge
only, on the other hand has tension $\tilde{T}_{(0,1)} = \lambda_B^{-1}
T_1^{(B)}$.  The scaling $T \sim \lambda^{-1}$ is characteristic of $D$-branes
in the string metric.\cite{polchinski95}  This is to be contrasted with ordinary solitons, like
the `t Hooft--Polyakov monopole, which have $T \sim \lambda^{-2}$.

As is typical of BPS mass formulas, the tensions we have found satisfy a
triangle inequality
\begin{equation}
T_{(p_{1} + q_{1}, p_{2} + q_{2})} \leq T_{(p_{1}, p_{2})} + T_{(q_{1},
q_{2})},
\end{equation}
and equality requires that $\vec p$ and $\vec q$ are parallel.  This means that
if $q_1$ and $q_2$ are relatively prime, a string with charges $(q_1, q_2)$ and
tension $T_{(q_{1}, q_{2})}$ is absolutely stable, protected by charge
conservation and a ``tension gap'' (the analog of a mass gap) from decay into
multiple strings.  On the other hand, a string with charges $(nq_1, nq_2)$ is
at the threshold for decay into $n (q_1, q_2)$ strings.  Whether one has a bound
state or not, in such a case, is a delicate issue whose answer depends on the
particular problem.  We will show that  the duality relation to M  theory
requires that only strings with $q_1$ and $q_2$ relatively prime be included.
This conclusion is supported by a bound-state analysis carried out by Witten.\cite{witten95b}
One way of stating the conclusion is that $q_1$ fundamental strings and $q_2$
$D$-strings (all of which are parallel) can form a single bound state if and
only if $q_1$ and $q_2$ are relatively prime.  It should also be noted that the
$(-q_1, -q_2)$ string is the orientation-reversed $(q_1, q_2)$ string.

\subsection{Compactification of IIB Theory on a Circle}

Let us now consider type IIB string theory compactified on a circle of radius
$R_B$ (and circumference $L_B = 2\pi R_B)$.  Since all of the $(q_1, q_2)$
strings are related by $SL(2,\ZZ)$ transformations, they are all equivalent and
any one of them can be weakly coupled.  However, when one is weakly coupled,
all the others are necessarily strongly coupled.  Nevertheless, let us consider
an arbitrary $(q_1, q_2)$ string and write down the spectrum of its 9d excitations
in the limit of weak coupling.  This is given by standard string theory
formulas:
\begin{equation}
M_B^2 = \left({m\over R_B}\right)^2 + (2\pi R_B n T_{(q_{1}, q_{2})})^2
 + 4\pi T_{(q_{1}, q_{2})} (N_L + N_R).   \label{twoc}
\end{equation}
Here $m$  is the Kaluza--Klein excitation number and $n$ is the string winding
number.  $N_L$ and $N_R$ are excitation numbers of left-moving and right-moving
oscillator modes, and the level-matching condition is
\begin{equation}
N_R - N_L = mn.
\end{equation}

Now our purpose is to use this formula for all the $(q_1, q_2)$ strings
simultaneously.  However, the formula is completely meaningless at
strong coupling, and (as we have said) at most one of the strings is weakly
coupled.  The appropriate trick in this case is to consider only BPS states - ones
belonging to short supersymmetry multiplets.  They are easy to identify,
being given by either $N_L = 0$ or $N_R = 0$.  (Ones with $N_L = N_R = 0$ are
ultrashort.)  For these states the mass formula should be exact, even at strong
coupling.  Therefore, it can be used for all the strings at the same time.  In this
way, we obtain reliable mass formulas for a very large part of the spectrum -- much more than
appears in perturbation theory.  Of course, the appearance of this rich
spectrum of BPS states depends crucially on the compactification.  

Using eqs.~(\ref{twoa}) and (\ref{twob}), the winding-mode term in eq.~(\ref{twoc}) 
contains the factor
\begin{equation}
n^2 \Delta_{(q_1,q_2)} = e^{\phi_{0}} |\ell_1 - \ell_2 \rho_0|^2,
\end{equation}
where $(\ell_1, \ell_2) = n (q_1, q_2)$.  There is a unique correspondence
between the three integers $n, q_1, q_2$ and an arbitrary pair of integers $\ell_1, \ell_2$.  The
integer $n$ is the greatest common division of $\ell_1$ and $\ell_2$.  The only
ambiguity is whether to choose $n$ or $-n$, but since $n$ is (oriented) winding
number and the $(-q_1, -q_2)$ string is the orientation-reversed $(q_1, q_2)$
string, the two choices are actually equivalent.  Thus BPS states are
characterized by three integers $m, \ell_1, \ell_2$ and oscillator excitations
corresponding to $N_L = |mn|$, tensored with a 16-dimensional short multiplet from
the $N_R=0$ sector (or vice versa).

\subsection{Comparison with M Theory on a Torus}

Let us now consider 11d M  theory compactified on a torus.  The torus is
characterized by a complex modulus $\tau = \tau_1 + i \tau_2$ (as usual) and by
its area ${A}_M$, measured in the 11d canonical metric.  If the two
periods of the torus are $2\pi R_{11}$ and $2\pi R_{11} \tau$, then $A_M =
(2\pi R_{11})^2 \tau_2$.  In terms of coordinates $z = x + i y$ on the torus,
a single-valued wave function has the form
\begin{equation}
\phi_{\ell_{1}, \ell_{2}} \sim {\rm exp} \left\{{i\over R_{11}} \left[x \ell_2 -
{y\over\tau_2} (\ell_2 \tau_1 - \ell_1)\right]\right\}.
\end{equation}
These characterize Kaluza--Klein excitations.  The contribution to the
mass-squared is given by the eigenvalue of $- \partial_x^2 - \partial_y^2$,
\begin{equation}
{1\over R_{11}^2} \left(\ell_2^2 + {1\over\tau_2^2} (\ell_2 \tau_1 -
\ell_1)^2\right) = {|\ell_1 - \ell_2 \tau|^2\over (\tau_2 R_{11})^2}.
\end{equation}
Our purpose is to match BPS states of M  theory on $T^2$ and IIB theory on
$S^1$.  Clearly, this term has the right structure to match the string
winding-mode terms described at the end of the last subsection, provided that
we make the identification\cite{schwarz95a,aspinwall95a}
\begin{equation}
\tau = \rho_0.
\end{equation}
The normalizations of the two terms we are matching are not the same, but that
is because they are measured in different metrics.  The matching tells us
how to relate the two metrics, a formula to be presented soon.  For now, let
me emphasize that the identification $\tau = \rho_0$ is a pleasant surprise,
because it implies that the non-perturbative $SL(2,\ZZ)$ symmetry of the IIB
theory, after compactification on a circle, can be reinterpreted as the modular
group of a toroidal compactification!  Of course, once the symmetry is
established for finite $R_B$, it should also persist in the limit $R_B
\rightarrow \infty$.

To go further, we also need an M  theory counterpart of the term $(m/R_B)^2$
in the IIB string mass formula.  Here there is also a natural candidate:
wrapping M  theory 2-branes so as to cover the torus M  times.  If
the 2-brane tension is $T_2^{(M)}$, this gives a contribution $(A_M
T_2^{(M)} m)^2$ to the mass-squared.  Matching the normalization of this term,
as well as the Kaluza--Klein term, one learns that the metrics are related by
\begin{equation}
g^{(M)} = \beta^2 g^{(B)}, \label{twod}
\end{equation}
where
\begin{equation}
\beta^2 = A_M^{1/2} T_2^{(M)}/T_1^{(B)},  \label{twoe}
\end{equation}
and that the compactification volumes are related by
\begin{equation}
(T_1^{(B)} L_B^2)^{-1} = {1\over (2\pi)^2} T_2^{(M)} A_M^{3/2}.  \label{twof}
\end{equation}
Since all the other factors are constants, this gives (for fixed $\tau = \rho_0$)
the scaling law $L_B \sim A_M^{-3/4}$.

We still have the oscillator excitations of the type IIB string BPS mass formula to
account for.  Their M  theory counterparts must be excitations of the wrapped
2-brane.  Unfortunately, since the quantization of the 2-brane is not
yet understood, this cannot be checked.  The story could be
turned around at this point to infer what the BPS excitation spectrum 
of wrapped 2-branes must be.
Maybe, trying to understand this spectrum will lead to a better understanding
of 2-brane quantization.  In any case, assuming that this works, we have found that
Kaluza--Klein excitations of the type IIB theory 
compactified on a circle correspond to wrappings
of the 2-brane on the torus and that Kaluza--Klein modes of M  theory on the
torus correspond to windings of an infinite family of type IIB strings on the
circle.

In the preceding discussion
it was not specified exactly how the M  theory 2-brane wraps on the torus
when it ``covers it $m$  times.''  There are a variety of different possible
maps that could define the mapping, and it should be specified which ones are
allowed and what is the proper way to count them.  I do not have a complete
answer to this question, but there is one comment that may prove useful.  In
the simpler problem of M  theory compactified on a circle, the IIA string in
10d arises from wrapping one cycle of a toroidal 2-brane on the spatial
circle.\cite{duff87}  If one were to wrap the circle $m$  times, instead, this would appear to
give a IIA string of $m$  times the usual tension.  However, it is quite clear
that no such string exists, so such a configuration must be unstable to decay
into $m$ strings.  Perhaps the rule is that one cycle should be wrapped only
once and the dual one $m$ times, but this requires identifying a preferred
cycle on the torus.  The only preferred cycle in the problem is the one defined
by the Kaluza--Klein excitation.

\subsection{Matching p-branes}

We have conjectured that M  theory compactified on a torus of area 
${A}_M$ and modular parameter $\tau$ is identical to type IIB string theory
compactified on a circle of circumference $L_B$ and vacuum parameter $\rho_0$.
The conjecture was supported by matching BPS 0-branes in 9d, which
dictated how to match parameters $(\tau = \rho_0,$ etc.).  
We can carry out additional tests of the proposed
duality, and learn interesting new relations at the same time,  by also
matching BPS $p$-branes with $p > 0$ in 9d.\cite{schwarz95b} 
Here we will describe the results
for $p = 1,2,3,4$, though other cases can also be analyzed.

Let us start with $p = 1$ (strings) in 9d.  Trivial reduction of the IIB strings
in 10d gives strings with the same charges $(q_1, q_2)$ and tensions
$T_{(q_{1}, q_{2})}$ in 9d.  The interesting question is how these should be
interpreted in M  theory.  The way to do this is to start with a 2-brane of toroidal
topology in M theory and to wrap one of its cycles on a $(q_1, q_2)$ homology cycle
of the spatial torus.  The minimal length of such a cycle is
\begin{equation}
L_{(q_{1}, q_{2})} = 2 \pi R_{11}|q_1 - q_2 \tau| = (A_M \Delta_{(q_1,q_2)})^{1/2}.  \label{twoi}
\end{equation}
Thus, this wrapping gives a 9d string whose tension is
\begin{equation}
T_{(q_{1}, q_{2})}^{(11)} = L_{(q_{1}, q_{2})} T_2^{(M)}.
\end{equation}
The superscript 11 emphasizes that this is measured in the 11d metric.  To
compare with the IIB string tensions, we use eqs.~(\ref{twod}) and (\ref{twoe}) to deduce that
\begin{equation}
T_{(q_{1}, q_{2})} = \beta^{-2} T_{(q_{1}, q_{2})}^{(11)} = (\Delta_{(q_{1},
q_{2})})^{1/2} T_1^{(B)}.
\end{equation}
This agrees with the result in subsection 4, showing that this is a correct
interpretation.

To match 2-branes in 9d we must wrap the IIB theory 3-brane on the circle and
compare to the M theory 2-brane.  The wrapped 3-brane gives a 2-brane with
tension $L_B T_3^{(B)}$.  Including the metric conversion factor, the matching
gives
\begin{equation}
T_2^{(M)} = \beta^3 L_B T_3^{(B)}.
\end{equation}
Combining this with eqs.~(\ref{twoe}) and (\ref{twof}) gives the identity
\begin{equation}
T_3^{(B)} = { 1\over 2\pi} (T_1^{(B)})^2. \label{twoh}
\end{equation}
It is remarkable that the M theory/IIB theory duality not only relates M
theory tensions to IIB theory tensions, but it implies a relation involving
only IIB tensions.  The 3-brane tension is a constant in the canonical metric, but 
using eq.~(\ref{twog}) it scales as $\lambda_B^{-1}$ in the string metric, as expected for a
$D$-brane.

Wrapping the M theory 5-brane on the spatial torus gives a 9d 3-brane, which
can be identified with the IIB theory 3-brane reduced to 9d.  This gives
\begin{equation}
T_5^{(M)} A_M  \beta^4 T_3^{(B)},
\end{equation}
which combined with eqs.~(\ref{twoe}) and (\ref{twoh}) implies that
\begin{equation}
T_5^{(M)} = {1\over 2\pi} (T_2^{(M)})^2.  \label{twoj}
\end{equation}
This corresponds to satisfying the Dirac quantization condition with the
minimum allowed product of charges.\footnote{According to Ref. ~\cite{duff95a}, it
corresponds to one-half of the minimum product. The result given here has been confirmed
by $D$-brane arguments, so I am quite sure it is correct.}

The matching of 4-branes works similarly.  The IIB theory has an infinite
family of 5-branes with tensions $T_{5(q_{1}, q_{2})}^{(B)}$.  Wrapping a cycle
on the spatial circle gives a family of 9d 4-branes with tensions $L_B
T_{5(q_{1}, q_{2})}^{(B)}$.  This should match the 4-branes obtained by
wrapping the M theory 5-brane on a $(q_1, q_2)$ homology cycle of the spatial
torus.  Thus,
\begin{equation}
T_5^{(M)} L_{(q_{1}, q_{2})} = \beta^5 L_B T_{5(q_{1}, q_{2})}^{(B)}.
\end{equation}
Combined with eqs.~(\ref{twoe}), (\ref{twof}), (\ref{twoi}), 
and (\ref{twoj}), this implies that the IIB 5-brane tensions are given by
\begin{equation}
T_{5(q_{1}, q_{2})}^{(B)} = {1\over (2\pi)^2} (\Delta_{(q_1, q_2)})^{1/2}
(T_1^{(B)})^3.
\end{equation}
In this case $q_1$ is the magnetic RR charge and $q_2$ is the magnetic
NS-NS charge.  Thus, the tension of a 5-brane with pure RR charge scales as
$\lambda_B^{1/2}$ and the tension of one with pure NS-NS charge scales as
$\lambda_B^{1/2}$.  Converting to the string metric, these become
$\lambda_B^{-1}$ and $\lambda_B^{-2}$, respectively, as expected for a $D$-brane
and an ordinary soliton.

The matching of 5-branes in 9d works differently.  The M theory 5-brane
reduced to 9d corresponds in the IIB picture to the Kaluza--Klein 5-brane,
which couples magnetically to the $U(1)$ gauge field associated to the isometry
of the circle.  Similarly, the two-parameter family of IIB theory 5-branes
corresponds to the two-parameter family of Kaluza--Klein 5-branes, in the M
theory description, which couple magnetically to the two $U(1)$ gauge fields associated
to the two isometries of the torus.

\subsection{Implications of the Duality}

What does the IIB/M theory duality mean?  Certain facts are an immediate
consequence of the scaling rule $L_B \sim A_M^{-3/4}$.  Namely, compactifying
M theory on a
torus and letting $A_M \rightarrow 0$, while holding $\tau$ fixed, gives the
IIB theory in 10d in the limit.  Similarly, compactifying the IIB theory on a
circle and letting $L_B \rightarrow 0$, for fixed modulus $\rho_0$, gives M
theory in 11d in the limit.

When $L_B$ and $A_M$ are finite, and the vacuum has 
only 9d Poincar\'e symmetry, one
might ask ``how many compactified dimensions are there?''  From the IIB
viewpoint there is one, and from the M theory viewpoint there is two.  Is one of these
answers better than the other?  Can they be combined and regarded as three
compact dimensions?  To see what is happening it is instructive to list the 9d
massless bosonic fields showing
their corresponding M theory and IIB descriptions.
\begin{equation}
\begin{array}{rl}
{\bf M \ theory} & \quad \quad {\bf IIB\  theory}\\
g_{\mu\nu}^{(M)} \quad  & \quad  \quad \quad g_{\mu\nu}^{(B)}\\
g_{\mu\alpha}^{(M)} \quad  & \quad  \quad \quad B_{\mu 9}^{(\alpha)}\\
g_{\alpha\beta}^{(M)} \quad  & \quad  \quad \quad \rho, g_{99}^{(B)}\\
C_{\mu\nu\rho} \quad  & \quad  \quad \quad A_{\mu\nu\rho 9}\\
C_{\mu\nu\alpha} \quad  & \quad  \quad \quad B_{\mu\nu}^{(\alpha)}\\
C_{\mu\nu\alpha\beta} \quad  & \quad  \quad \quad g_{\mu 9}^{(B)}\\
\end{array} \label{lista}
\end{equation}
The indices $\alpha, \beta = 1,2$ refer to the two internal
directions of the M theory torus, the index 9 refers to the IIB theory circle,
and $\mu, \nu$ are 9d indices.  What the list demonstrates is that which
fields are ``matter'' and which ones are ``geometrical'' is subjective,
depending on whether you adopt an M theory or IIB theory viewpoint.  Both
viewpoints are valid, and neither is preferable to the other.  So, how many
compact dimensions there are is just a matter of how the fields are labelled!
However, there is no straightforward choice of labelling that exhibits three
compact dimensions.  In my opinion, some of the recent suggestions that these
theories can be derived from 12d are effectively counting the dimensions of both the
torus and the circle.

The tests and implications of the M theory/IIB theory duality that we have
presented so far are certainly not the only ones.  For example,  there are other
solitons -- intersecting $p$-branes, for example -- that break 3/4 or 7/8 of the 
supersymmetry.\cite{papadopoulos96}  
They still have good BPS saturation properties, so that they are under control. 
It would be instructive to consider the matching of these solitons in 9d, too,
something that has not yet been done.

\subsection{M Theory/SO(32) Theory Duality}

As we mentioned at the beginning of this section, there is a second duality
that is closely related (and, therefore, quite similar) to the one we have been
discussing.  It relates M theory compactified on $S^1/ \ZZ_2 \times S^1$ to
$SO(32)$ theory compactified on $S^1$.\cite{schwarz96} 
Since $S^1/ \ZZ_2$ can be regarded as a
line interval $I,$ $S^1/ \ZZ_2 \times S^1$ can be regarded as a cylinder $C$.  We
will choose its height to be $L_1$ and its circumference to be $L_2 = 2\pi
R_2$.   The circumference of the circle on which the $SO(32)$ theory is
compactified is denoted $L_O = 2\pi R_O$ in the canonical 10d metric.

Before describing $p$-brane matching in 9d, let us briefly review the
Horava--Witten picture of the $E_8 \times E_8$ heterotic string theory.\cite{horava95}
Compactification of M theory on $S^1/\ZZ_2 = I$ gives a
space-time with two 10d faces, separated by a distance $L_1 \sim
\lambda_H^{2/3}$, where $\lambda_H$ is the coupling constant of the $E_8 \times
E_8$ theory in 10d.  The two 10d faces are sometimes called ``end-of-the-world
9-branes.''  Each of them carries the gauge fields for one of the two $E_8$'s.
For reasons that will be explained in the next section, M theory 2-branes are
allowed to terminate on a face, so that the boundary of the 2-brane is a circle
inside the face.  In this picture, an $E_8 \times E_8$ heterotic string is a
cylindrical 2-brane suspended between the two faces, with one $E_8$ current
algebra associated to each boundary.  This cylinder (or strip) is well
approximated by a string living in 10d when the separation $L_1$ is small.
Since perturbation theory in $\lambda_H$ is an expansion about $L_1 = 0$, the
fact that there really are eleven dimensions and that the string 
is actually a membrane is invisible in that approach.

The story described above is very similar to the relation between the IIA
superstring and M theory.  In that case the compact dimension is a circle and
the IIA string arises from wrapping the M theory 2-brane around the circle.
Thus, in a sense, the non-perturbative $E_8 \times E_8$ theory just involves
modding out the non-perturbative IIA theory by a $\ZZ_2$.  (This is a bit glib,
since the rules for carrying out the modding in M theory, which
is not a string theory, are not so obvious.)
Similarly, the type I $SO(32)$ string theory in 10d can be constructed as a
$\ZZ_2$ orientifold of the type IIB theory in 10d.  Thus, the duality we are
considering now can be viewed as arising from modding out the previous one by a
$\ZZ_2$ on both sides.  However, in the following we will treat it as a
separate problem instead of attempting to exploit that picture.  Because of the
similarity of the two problems, fewer details will be provided this time.

The $SO(32)$ theory in 10d has both a type I and a heterotic description, which
are $S$ dual.  That is, their coupling constants, $\lambda = e^{<\phi>}$,
satisfy $\lambda_H^{(0)} = (\lambda_I^{(0)})^{-1}$.  As before, we match
supersymmetry-preserving (BPS) branes in 9d.  Recall that in the $SO(32)$
theory, there is just one two-form field $B_{\mu\nu}$, and the $p$-branes that
couple to it are the $SO(32)$ heterotic string and its magnetic dual, which is
a 5-brane.  The type I open and closed strings do not carry a conserved charge
and are not BPS.  This is the reason that they can break.  So from the type I
viewpoint it is clear that the heterotic string can give  a $0$-brane or a $1$-brane
in 9d and that the dual 5-brane can give a 5-brane or a 4-brane in 9d.  In each case,
the issue is simply whether or not one cycle wraps around the spatial circle.

Now we need to find the corresponding 9d $p$-branes from the M theory
viewpoint, to understand why they are the only ones, and to explore what can be
learned from matching tensions.  We described how the $E_8 \times E_8$ string
arises in 10d from wrapping the M theory 2-brane on $I$.  Subsequent
reduction on a circle can clearly give a $0$-brane or a $1$-brane.  But why is
there no BPS 2-brane in 9d?  When the 2-brane is forced to be in a 10d boundary,
rather than in the 11d bulk, it becomes breakable (non-BPS).   The technical
reason (see the next section) is that there is no 3-form gauge field on the
boundary (or in the 10d reduction).  The story for the five-brane is just the
reverse.  Whereas the 2-brane must wrap on the $I$ dimension, the five-brane must
not do so.  As a result it gives a 5-brane or a 4-brane in 9d according to
whether or not it wraps around the $S^1$ dimension.  So, altogether, both
pictures give the electric-magnetic dual pairs ($0,5$) and ($1,4$) in 9d.

{}From the $p$-brane matching one learns that
\begin{equation}
\lambda_H^{(0)} = {L_1\over L_2}.
\end{equation}
Thus, the $SO(32)$ heterotic string is weakly coupled when the spatial cylinder
of the M theory compactification is a thin ribbon $(L_1 \ll L_2)$.  This is
consistent with the earlier conclusion that the $E_8 \times E_8$ heterotic string
is weakly coupled when $L_1$ is small.  Conversely, the type I string is weakly
coupled for $L_2 \ll L_1$, in which case the spatial cylinder is long and thin.
 The $\ZZ_2$ transformation that inverts the modulus of the cylinder,
$L_1/L_2$, corresponds to the type I/heterotic duality of the $SO(32)$ theory.
Since it is not a symmetry of the cylinder it implies that two
different-looking string theories are $S$ dual.  This is to be contrasted with
the $SL(2,\ZZ)$ modular group symmetry of the torus, which accounts for the
self-duality of the IIB theory.

The $p$-brane matching in 9d gives the relation
\begin{equation}
L_1 L_2^2 T_2^{(M)} = \left({T_1^{(0)} L_0^2\over 2\pi}\right)^{-1},
\end{equation}
which is the analog of eq.~\ref{twof}.  As in that case, it tells us that for fixed
modulus $L_1/L_2$, one has the scaling law $L_O \sim A_C^{-3/4}$, where $A_C =
L_1 L_2$ is the area of the cylinder.  Eq.~\ref{twoj} relating $T_2^{(M)}$ and
$T_5^{(M)}$ is reobtained, and one also learns that
\begin{equation}
T_5^{(O)} = {1\over (2\pi)^2} \left({L_2\over L_1}\right)^2 (T_1^{(O)})^3.
\end{equation}
In the heterotic string metric, where $T_1^{(O)}$ is a constant, this implies
that $T_5^{(O)} \sim (\lambda_H^{(O)})^{-2}$, as is typical of a soliton.  In
the type I superstring metric, on the other hand it implies that $T_1^{(O)}
\sim 1/\lambda_I^{(O)}$ and $T_5^{(O)} \sim 1/\lambda_I^{(O)}$, consistent with
the fact that both are $D$-branes from the type I viewpoint.  So, what kind of
object you have depends very much on your point of view.

\subsection{Some Remarks on the Origins of Chirality in M Theory}

In its heyday (around 1980) there were two major reasons for being skeptical about
11d supergravity.  The first was its evident lack of renormalizability, which
led to the belief that it does not approximate a well-defined quantum
theory.  The second was its lack of chirality (\ie, its left-right symmetry),
which suggested that it could not have a vacuum with the chiral structure
required for a realistic model.  Our attitude towards both these issues now
needs to be reconsidered.  First, we now view 11d supergravity as a low-energy
effective description of M theory.  As such, it seems reasonable
to believe that there is a well-defined quantum interpretation.  The
situation with regard to chirality is also changed.  
Here the new ingredients are the branes -- the
2-brane and 5-brane, as well as the end-of-the-world 9-branes.  They can and do
introduce left-right asymmetry (consistent with anomaly cancellation requirements).

In the duality between M theory on a torus and IIB theory on a circle, that
we have been discussing, the issue of chirality already appears.  In the limit
that the area $A_M$ of the torus vanishes, one obtains the IIB theory in 10d,
which is a chiral theory.  Let's track down the M theory origins
of the chiral asymmetry.  One
question is whether it is property of the limit $A_M \rightarrow 0$, or whether
it is already visible in 9d.  Let's look at this question first from the IIB
viewpoint.

The chiral fields of the IIB theory in 10d are the massless fermions and the
four-form $A_4$, whose field strength is self-dual.  This means that the
associated physical degrees of freedom belong to parity non-invariant
representations of the massless little group, spin (8).  Compactifying on a
circle, they give BPS Kaluza--Klein towers of excitations with $M_n^2 =
(n/R_B)^2$.  These belong now to representations of the massive 9d little group, which
is also spin (8).  Indeed, it is obvious that these are the same parity
non-invariant representations we started with.  So, in this sense, one could
say that these massive excitations are chiral.  Certainly, they account for the
chirality of the massless 10d field in the decompactification limit.  The
massive 4-form modes in 9d are described by complex fields (combining $n$ and
$-n$).  Dropping the index $n$, they satisfy a free wave equation of the form
$A_4 \sim im * F_5$.\cite{townsend84a}  
Taking the exterior derivative, $F_5 \sim imd (*F_5)$.
Even though there is no manifestly covariant action in 10d for a 4-form with a
self-dual field strength, there is one for this massive complex 4-form in 9d:
\begin{equation}
{\cal L} (A_4) \sim \int F_5^* \cdot F_5 d^9 x + im \int A_4^* \wedge F_5.
\end{equation}
As expected, this has a parity-violating mass term.  The number of propagating
modes described by such a Lagrangian is the same for $m \not= 0$ as it is for
$m = 0$.  This structure is quite similar to the much-studied ``topologically
massive gauge theory'' in 3d: \cite{deser82}
\begin{equation}
{\cal L} (A_1) \sim \int F_2 \cdot F_2 d^3 x + m \int A_1 \wedge F_2.
\end{equation}
One difference is that the construction can be carried out for real fields in
dimensions $4k - 1$ whereas complex fields are required in dimensions $4k + 1$.

Having identified where the chirality resides in 9d, we can now ask how 
these states originate in the 11d description.  The answer is immediate, because
one of the things we learned from studying the M theory/IIB theory duality is
that type IIB  Kaluza--Klein excitations on $S^1$ correspond to M theory
wrapping modes of the 2-brane on $T^2$.  Therefore, the massive chiral modes in
9d must arise from wrapping the 2-brane.  The 2-brane world-volume theory
itself is 3d, so one might think it could not be chiral.  However,
it contains a ``charge'' coupling to the background 3-form gauge field $\int
C_{\mu\nu\rho} dx^\mu \wedge dx^\nu \wedge dx^\rho$, and this is precisely the relevant
chiral term.  The 3-form gives a $U(1)$ gauge field in 9d with two indices in
internal directions $(C_{\mu 12})$, and the massive chiral modes due to $n$ units
of wrapping carry $n$ units of electric charge, as measured by this gauge field in
9d.  The corresponding field in the IIB picture is the Kaluza--Klein gauge
field $g_{\mu 9}$.  (This correspondence already appeared in eq.~(\ref{lista}).)

\section{WHICH BRANES CAN END ON WHICH?}

In considering the Horava--Witten description of the $E_8 \times E_8$ heterotic string,
we concluded that in M theory  a 2-brane can terminate on an
end-of-the-world 9-brane.  This section discusses, with
several examples, the general question of
when one supersymmetric (BPS) brane is allowed to
terminate on another one.  Following Strominger, we argue that charge
conservation is an essential consideration.\cite{strominger95a} 
However, we will discover that there is also a subtle ``wormhole'' construction,
which gives additional possibilities.  The basic idea is explained by
$D$-branes, which are defined as $p$-branes on which  strings can
terminate.\cite{polchinski95}  
The type II (A or B) fundamental string carries a conserved charge
and couples electrically to the NS-NS sector two-form gauge field $B_{\mu\nu}$.
When a string carrying this kind of conserved charge has
an end, flux associated with a $U(1)$ gauge field 
emerges from the end into the $D$-brane. This means that there is a point electric charge on
the end of the string (explaining Chan--Paton factors),
and that the $D$-brane world-volume theory
contains a $U(1)$ gauge field $A_\alpha$, which can assume a suitable
configuration to carry away the flux.  An interesting generalization
is the case of $N$ coincident $D$-branes when the
$[U(1)]^N$ gauge symmetry of the individual $D$-branes gets extended to a 
non-Abelian $U(N)$
gauge symmetry.  Let us turn now to specific examples of branes ending on
branes.

\subsection{Three Examples}

An example of a $D$-brane is the 2-brane of type IIA string theory in 
10d.  As with all type II $D$-branes, massless fields of the world-volume
theory form an $N=1$ 10d gauge multiplet, restricted to the brane.  The
vector $A_\mu$ of the 10d gauge theory decomposes into a three-vector
$A_\alpha$ and seven scalars $\phi_i$ in the world-volume theory.  The scalars
can be regarded as collective coordinates for excitations of the brane in the
seven transverse dimensions or as Goldstone bosons for
broken translational symmetries. Similarly, the world-sheet fermions
corrspond to broken supersymmetries.  The gauge field $A_\alpha$ carries
off the flux when a type IIA string terminates on the 2-brane.
{}From the point of view of this gauge field, the charge on the end of the
string is electric.  However, there is a dual magnetic picture, which is also
interesting.  The world-volume theory of the 2-brane can be recast by a
duality transformation $(dA = * d \phi_8)$ that replaces $A_\alpha$ by an
eighth scalar $\phi_8$.\cite{duff93,townsend95c}  
This scalar is a zero-form gauge field, and from its
point of view the charge on the end of the string is magnetic.  However, a more
profound viewpoint is that $\phi_8$ represents excitations in an eighth
transverse dimension, so that the 2-brane actually lives in an
11d space-time.  This strongly suggests that, after the duality transformation, 
one is describing the 2-brane of M theory.  But this raises a paradox:
M theory in 11d Minkowski space does not have strings that can
terminate on the 2-brane, so what is the strong coupling description of a
configuration consisting of a type IIA string ending a 2-brane?  I'll return
to this question later in this section.

Let's now consider the Dirichlet 5-brane of the type IIB theory.  By the same
reasoning as before, the 6d world-volume theory in this case contains a
six-vector $A_\alpha$ and four scalars $\phi_i$ representing transverse
excitations.  Let us once again replace the $U(1)$ gauge field $A_\alpha$ by a
dual gauge field.  In 6d the dual gauge field is a three-form.  As explained in
Section 2.2, a 2-brane can couple electrically to a three-form.  Thus, BPS
2-branes can live inside the 5-brane.  What this means is that the 3-brane
of the IIB theory can terminate on the 5-brane.\cite{strominger95a}  
Its boundary is a 2-brane, and the charge that exists on its boundary gives rise to electric
flux of the three-form gauge field of the 5-brane.  Equivalently, had we not
made a duality transformation, it would give magnetic flux of the original
$U(1)$ gauge field.  Thus, it is consistent with charge conservation for a type IIB
3-brane to terminate on a Dirichlet 5-brane.  The $SL(2,\ZZ)$ duality
symmetry of the type IIB theory can be invoked to draw additional conclusions.
 Under an $SL(2,\ZZ)$ transformation the 3-brane is invariant, but the
Dirichlet 5-brane, which carries $B_{\mu\nu}^{(I)}$ magnetic charges $(0,1)$,
can be transformed into a $(q_1, q_2)$ 5-brane.  This implies that the
3-brane is allowed to terminate on any of the 5-branes.

As a third example, let us consider the M theory 5-brane.  Its massless sector
consists of a 6d $N=2$ tensor supermultiplet.\cite{callan91,becker96}  The bosons in this
multiplet are a two-form gauge field $B_{\alpha\beta}$, with a self-dual field
strength $(dB = - * dB)$, and five scalars $\phi_i$ describing transverse
excitations of the 5-brane in 11d.  The fact that this is the
appropriate multiplet can be argued in many different ways.  Here we will
simply remark that this is the only matter supermultiplet with the correct
supersymmetry, and it contains the desired number of scalar fields.  The
two-form can couple to a self-dual string, which can be identified as the
boundary of 2-brane that ends on the 5-brane.  Thus, the M theory
2-brane can terminate on the M theory 5-brane, but not on another
2-brane.  Thus, the M theory 5-brane can be regarded as a
higher-dimension analog of a $D$-brane.  Rather than being defined as an object on
which an open string can end, it is an object on which an open membrane can
terminate.  The reason that the $D$-brane picture is so powerful is
that open strings can be quantized, and they can be used to describe
excitations of the $D$-brane.  We do not have this kind of mathematical control
for open membranes, so the $D$-brane picture is less useful (at the present time)
in the case of M
theory.  However, compactification on a circle makes it clear that this is
really more than an analogy.  If the compactification is arranged so that one
dimension of both the 2-brane and the 5-brane are wrapped on the spatial
circle, then the resulting 10d picture precisely corresponds to 
a IIA string ending on a IIA 4-brane. This 4-brane is a standard $D$-brane.

\subsection{Parallel p-branes}

When a $p_1$-brane is allowed to end on a $p_2$-brane, then it is also possible
to consider a pair of parallel $p_2$ branes with an open $p_1$-brane suspended
between them.  To be clear what we are talking about, let me emphasize that all
the branes under consideration are supersymmetric (BPS) branes carrying
conserved charges.  This means (for $p_2 < D-3$) that a pair of parallel
$p_2$-branes (infinite hyperplanes) is a stable configuration because the
forces between the branes cancel in such a case.  If one imagines attaching an open
$p_1$-brane that connects them, its tension would cause some
bending of the $p_2$-branes in the vicinity of the junction.  As far as I know,
explicit field configurations that realize this picture have not
been studied, but the qualitative picture is clear. In any case, the main reason to be interested
in such configurations is as a way of thinking about quantum excitations of a
system of parallel $p$-branes. New classical configurations are more of a curiosity.

As the separation of two $p_2$-branes becomes small, the ``length'' $\ell$ of
the $p_1$-brane becomes small.  In this case, it can become a good
approximation to view the pair of $p_2$-branes 
(and the intervening space) as a single $p_2$-dimensional system
and the collapsed $p_1$-brane as a $(p_1-1)$-brane 
of tension $T_{NC} = \ell \, T_{p_{1}}$
inside this $p_2$-dimensional space.  The subscript NC denotes
`non-critical', since the tension of such branes can be arbitrarily small (as
$\ell \rightarrow 0$).  Note that in the case of the $E_8 \times E_8$ heterotic string 
the vanishing of the
tension as $\ell \rightarrow 0$ can be compensated by a Weyl rescaling of the
metric so that there is a finite tension in the limit.  
In other cases, such as the ones to be discussed here, such a
rescaling is not appropriate, and the non-critical $(p_1-1)$-brane has a
spectrum of excitations that become massless as $\ell \rightarrow 0$.  These
excitations should be viewed as possible excitations of a pair of nearly
coincident $p_2$-branes.

A word of warning is in order here. The solitons that are being described as
``$p$ dimensional'' are given by field configurations that spread to some
extent in the transverse dimensions. As explained to me by Maldacena, in certain
cases the limit in which the length $\ell \rightarrow 0$ can result in the size of
transverse spread becoming large at the same time. Then the simple geometric picture
becomes misleading. Most of the considerations that follow do not require taking
a limit $\ell \rightarrow 0$, so they are not subject to this criticism. The limit only
appears when one wants to identify non-critical branes.

The specific example that we will focus on here is a pair of parallel
3-branes in 9d.  Our purpose in doing this is to extract
additional implications of the duality between M theory compactified on a
torus and type IIB theory compactified on a circle.  We have already seen that
a single 3-brane in 9d can be viewed equally as a IIB 3-brane or as an M-theory 
5-brane wrapped on the spatial torus.  Now we wish to extend this
picture to a pair of parallel 3-branes including the possibility of
suspending other branes between them.  The plan is to first consider parallel
3-branes of the IIB theory and then parallel 5-branes of M theory.

\subsection{Parallel 3-branes of Type IIB Theory}

Since 3-branes of IIB theory are $D$-branes,  fundamental $(1,0)$ type
IIB strings can end on them.  Also, such a string can be suspended between a
pair of parallel 3-branes.  The 4d world-volume theory of this system is a
$U(2)$ gauge theory, spontaneously broken to $U(1) \times U(1)$ when the
separation $\ell > 0$.  The open string introduces a unit of electric charge
for a $U(1)$ subgroup.  (The appropriate $U(1)$ is the one inside the $SU(2)$
factor.)  The sign of the charge is tied to the orientation of the string.
An $SL(2,\ZZ)$ duality transformation gives the same configuration
with the fundamental $(1,0)$ string replaced by a $(q_1, q_2)$ string.  In this
case the lightest modes have mass
\begin{equation}
M = \ell \, T_{(q_{1}, q_{2})} = \ell (q_1^2 + \lambda_B^{-2} q_2^2)^{1/2} T_1^{(B)},
\end{equation}
where we have used the results of Section 2.4  for the tension of $(q_1, q_2)$ strings.  This
formula, which can be generalized to include a $\theta$ angle, 
agrees with the BPS formula for dyons of the $N = 4$ gauge
theory in 4d.  Thus, this picture relates 
the conjectured $SL(2,\ZZ)$ duality of $N = 4$ gauge theory in 4d
to that of type IIB superstring theory in 10d.  Note that
in this example it is $0$-branes that are becoming massless as $\ell
\rightarrow 0$.  This implies that only states with spin $J\leq 1$ have $M
\rightarrow 0$ as $\ell \rightarrow 0$.  This picture is also applicable in 9d
if a dimension orthogonal to the brane is compactified.

\subsection{Parallel 5-branes of M Theory}

As we have explained, an M theory 2-brane can end on a 5-brane, and,
therefore, a 2-brane can be suspended between a pair of parallel 5-branes.
 To make contact with the results in preceding subsection, we wish to take this
configuration and compactify on a spatial torus in such a way that we end up
with a string suspended between 3-branes in 9d.  The
interesting point is that this can be done such that the string is a $(q_1,
q_2)$ string.  This is to be expected because the torus is responsible for the
$SL(2,\ZZ)$ duality.

To give a specific realization suppose that the 5-brane spatial coordinates
are in the $x_2, x_3, x_4, x_5, x_6$ hyperplane, and the other coordinates are
fixed at two sets of values on the two 5-branes.  Now suppose that the $x_2$
and $x_3$ coordinates are compactified to form the spatial torus.  This implies
that the 5-branes are wrapped on the torus, giving parallel 3-branes in
9d.  Now consider a 2-brane connecting the two 5-branes.  To
get a string in 9d we want only one of its coordinates to be wrapped.
Therefore, suppose it lies in a plane defined by $x_1, x_2'$, with the other
coordinates fixed.  Here $x_2'$ is a line in the $x_2$ - $x_3$ plane, and hence on
the spatial torus.  It belongs to the $(q_1, q_2)$ homology class
if it satisfies $q_2 x_2 = q_1 x_3$.  In this way, we obtain what we wanted, a
$(q_1, q_2)$ string suspended between parallel 3-branes in 9d.  This is the
dual description of the configuration obtained in the preceding subsection

\subsection{A Paradox and its Resolution}

In the preceding subsection we considered an M theory 2-brane suspended
between parallel 5-branes and showed that it could be wrapped on a spatial
torus to give a $(q_1, q_2)$ string suspended between parallel 3-branes.
An alternative possibility would be to wrap two dimensions of the 5-branes
and no dimensions of the 2-brane on the spatial torus.  This gives a
2-brane suspended between parallel 3-branes in 9d.  What is
the IIB interpretation of this configuration?  The only simple guess is an
open 3-brane suspended between parallel 3-branes.  This is not allowed,
however, because the boundary of a 3-brane is a 2-brane and the world-volume
theory does not contain the gauge field needed to carry off its flux.  Before
explaining the right answer, let us first examine an analogous problem for
which the answer is known.

The IIA theory in 10d can have a string suspended between a pair of parallel
4-branes, since they are $D$-branes.  As we remarked earlier, this is
interpreted in M theory as a wrapping of an 11d configuration consisting of a
2-brane suspended between parallel 5-branes.  But what is the M theory
interpretation of a string suspended between parallel 2-branes?  The two problems
seem quite analogous, but the interpretations are rather different.  The simplest guess,
an open 2-brane suspended between parallel 2-branes is not allowed by
charge conservation, so that is not the answer.  The correct answer, which we
will now describe, was presented by Aharony, Sonnenschein, and 
Yankielowicz.\cite{aharony96}. (I  first heard it from Polchinski. Later, Yankielowicz 
explained it to me in detail and drew my attention to ref.~\cite{aharony96}.)

Consider an M theory 2-brane in a ``wormhole'' configuration.  This means
one smooth surface consisting of parallel 2-branes connected by a throat .  
As described, there are actually two {\it anti-parallel} branes connected
by a throat.  Such a configuration is highly unstable.  The size of the throat
would grow rapidly eating up the branes.  To stabilize the configuration one
must flip-over one of the two faces to make them parallel.  This can be
achieved by rotating one of them by $\pi$ in a plane orthogonal to the throat
-- the $x_3$ - $x_4$ plane, say.  This rotation involves no self-intersections of
the surface, but since more than three spatial dimensions are required, it is
somewhat difficult to visualize.  In any case, after doing this we have two
{\it parallel} branes connected by a throat.  This configuration is consistent
with charge conservation and represents an excitation of a BPS system.

Now the desired 10d configuration -- an open string suspended between parallel
2-branes -- can be obtained by compactification on a circle.  The geometry is
a bit subtle, however.  Suppose $y$ is the compact coordinate of the spatial
circle, and $x_2$ and $x_3$ are the coordinates that parametrize one of the faces.
Now imagine going around a circle on the face that encloses the throat.  By
continuity with the connecting tube (the throat), which is wrapped around the
$y$ direction, it is clear that this circle also winds around the $y$ direction.
The coordinate $y$ corresponds to the scalar field of the M theory 2-brane
world volume, which was called $\phi_8$ earlier when we introduced
it as a zero-form dual of a $U(1)$
gauge field.  The throat is a source of $y$ (or $\phi_8$) magnetic charge on
the end faces, since $\oint dy \not= 0$ when the contour encloses the throat.
In the dual formulation, appropriate to the IIA description of the membrane,
$\phi_8$ is replaced by a $U(1)$ gauge field $A_\alpha$ and this magnetic
charge is re-interpreted as electric charge.  This matches the picture expected
from the $D$-brane viewpoint, showing that we have found the correct M theory
interpretation.

Now it is clear what the answer to the original problem -- the IIB
interpretation of a 2-brane suspended between parallel 3-branes in 9d --
should be.  The correct 10d picture is a parallel 3-branes connected by a
throat, altogether forming one smooth surface.  Again one of the 3-branes
must be flipped over to ensure that they are parallel and not anti-parallel.
Compactification on a circle is then arranged so that one dimension of the
connecting throat is wrapped leaving a 2-brane from the 9d viewpoint.  The
throat again introduces a magnetic charge $\oint dy$.  In 9d this is best
viewed in the dual picture in which $y$ is replaced by a two-form, since the
3-brane has a 4d world volume.  A two-form is just what is needed to carry
off the flux associated to the charge on the string-like boundary of a 2-brane.
So, again, there is a consistent picture.

It was emphasized earlier that the $N=4$ gauge theory that describes the
world-volume theory of a pair of parallel 3-branes has an infinite spectrum of
dyons with $J\leq 1$ whose mass vanishes as $\ell \rightarrow 0$.  They were
interpreted in terms of strings suspended between the 3-branes.  Now we have
seen that in 9d there can also be a suspended 2-brane, which gives a noncritical string
as $\ell \rightarrow 0$.  The string modes describe states of
arbitrarily high spin whose mass vanishes as $\ell \rightarrow 0$ in the
effective $N = 4$ 4d gauge theory.

\subsection{Three-String Junctions}

The definition of a $D$-brane as a $p$-brane on which a type II string
can end has to be interpreted carefully for $p \leq 1$.  For example, in the
case of IIB strings, we found that there is an infinite family of strings with
$B_{\mu\nu}$ charges $(q_1, q_2)$, where $q_1$ and $q_2$ are relatively prime
integers.  The $(1,0)$ string is the fundamental string and the $(0,1)$ string
is the $D$-string.  A naive interpretation of `a fundamental string ending on a $D$-string' 
is described as the junction of three string segments, one of which is
$(1,0)$ and two of which are $(0,1)$.  This is not correct, however, because the
charge on the end of the fundamental string results in flux that must go into
one or the other of the attached string segments changing the string charge in
the process.  In short, the three-string junction must satisfy charge
conservation.\cite{aharony96}  
This means that an allowed junction that describes the joining of
three strings has charges $(q_1^{(i)}, q_2^{(i)})$, $ i = 1,2,3$
such that $q_1^{(i)}$ and $q_2^{(i)}$ are relatively prime for each value of
$i$ and $\sum_i q_1^{(i)} = \sum_i q_2^{(i)} = 0$.  The configuration is stable
if the three strings are semi-infinite and
the angles are chosen so that tensions, treated as vectors, add to zero.  It
would be interesting to construct the corresponding solution of the
supergravity field equations.  I expect it to be supersymmetric and, therefore, to
have the usual nice BPS properties.  For example, a periodic array could be
formed so that compactification on a circle would give the same configuration in
9d.

Given the three-string junction in 9d, it is natural to ask about its 11d M theory 
interpretation.  The answer is easily found and very pleasing.  In 11d
one could consider a three-membrane junction which consists of a single smooth
surface.  Topologically it is the same as the ``pants diagram,'' which describes
the world sheet for a closed string breaking into two closed strings.  Of
course, in the present problem the time coordinate is suppressed and the
surface is just a spatial diagram (like real pants).

Now we want to compactify on a torus to 9d in such a way that one
dimension of each of the three protruding 2-branes is wrapped on the torus.  Labelling
the three 2-branes by an index $i = 1,2,3$, we want the circular coordinate
 of the $i$th 2-brane to
wrap on a $(q_1^{(i)}, q_2^{(i)})$ homology cycle of the torus.  It is a simple
geometrical fact that this is only possible if $\sum_i q_1^{(i)} = \sum_i
q_2^{(i)} = 0$.  Thus, we reproduce the three-string junction in 9d
with exactly the desired properties.

This three-string junction could prove to be useful in future studies.  For
example, one might wish to consider a collection of wrapped, intersecting
3-branes as part of a study of black holes in string theory.  Excitations of
such a system would be represented by open strings connecting the 3-branes.
However, it would also be possible to have open strings ending on three different
3-branes if they are joined by a junction of the kind we have described.
Of course, given the existence of the three-string junction, one can also build
up more complicated networks, similar to $\phi^3$ Feynman diagrams.
 
\section{6D STRING VACUA WITH EXTENDED SUPERSYMMETRY}

Compactification to 9d was sufficient
to demonstrate that there are dualities connecting all superstring theories to
one another as well as to M theory.  Also, we learned that the dualities in 9d provide
non-perturbative information about these theories. 
In this section we will survey many (possibly all)
superstring vacua with extended supersymmetry in 6d.  
The next section will give a much less complete survey of 6d vacua with
N=1 supersymmetry. In the process, 
many new dualities that give additional insights into the
structure of the theory will appear.

Since vacua are characterized by their unbroken supersymmetry and massless
spectrum, among other things,  we need to know what the possibilities are in 6d.  Massless
particles are specified by representations of the little group, which is spin (4) or
$SU(2) \times SU(2)$.  Labelling representations by $SU(2) \times SU(2)$
multiplicities, one can have the following massless particles
\begin{equation}
\begin{array}{rl}
{\rm graviton}: & (3,3) \nonumber \\
{\rm gravitino}: & 2(3,2)\  {\rm or} \ 2(2,3)\nonumber \\
{\rm self \ dual\ tensor}: & (3,1)\ {\rm or}\ (1,3) \nonumber \\
{\rm vector}: & (2,2) \nonumber \\
{\rm spinor}: & 2(2,1)\ {\rm or}\  2(1,2)\nonumber \\
{\rm scalar}: & (1,1).
\end{array}
\end{equation}

Parity interchanges the two $SU(2)$'s.  Thus, the fermions and self-dual tensors are
chiral, while the other particles are not.  The factors of two in the fermion
multiplicities appear because a Weyl spinor in 6d is necessarily
complex.  The supersymmetry type can be labelled by the number of gravitinos of each
chirality.  The maximal case (32 supercharges) corresponds to $(2,2)$ or $N = 4$
and is non-chiral.  There are two $N = 2$ possibilities, just as in ten
dimensions: $(1,1)$ or IIA is non-chiral and $(2,0)$ or IIB is chiral.  $N = 1$
or $(1,0)$ supersymmetry is also chiral.  One might also consider
$(4,0), (3,1), (3,0)$, and $(2,1)$ supersymmetries, but I believe that none of
these is possible.\footnote{A gravity supermultiplet has been proposed for the $(2,1)$
case,\cite{strathdee87} but it gives gravitational anomalies 
that cannot be cancelled.}

\subsection{Vacua with (2,2) Supersymmetry}

Let us begin with maximal $(2,2)$ unbroken supersymmetry.  In this case
there is a unique massless multiplet -- the supergravity multiplet.  One way to determine
its particle is by decomposing the massless fields of 11d
supergravity into 6d pieces.  The result is as follows:
\begin{equation}
\begin{array}{rl}
{\rm bosons:} &(3,3) \!+ \!5(3,1) \!+ \!5(1,3) \!+ \!16(2,2) \!+
\!25(1,1)\nonumber \\
{\rm fermions:} &4(3,2) + 4(2,3) + 10(2,1) + 10(1,2).
\end{array}
\end{equation}
The $U$ duality group in this case is $SO(5,5;\ZZ)$ and the moduli space,
parametrized by the 25 scalar fields, is ${\cal M}_{5,5}$ .

A string vacuum with this amount of supersymmetry and this massless sector can
be obtained either by compactifying the type IIB theory on $T^4$ or M theory
on $T^5$.  Each construction explains a subgroup of the $U$ duality group.  The
type IIB theory compactification gives a $T$ duality group $SO(4,4;\ZZ)$ from the
$T^4$;  in the M theory compactification, the modular group of $T^5$
accounts for an $SL(5,\ZZ)$.  The complete answer, $SO(5,5;\ZZ)$, is
the smallest group that contains both of these as subgroups.  Once again, we see
the power of M theory/IIB theory duality.  By considering both at the same time
one can make deductions that are not apparent from either viewpoint separately.  The
multiplicities of the other boson fields are given by $SO(5,5)$
representations.  Thus, for example, the 16 vectors belong to a spinor representation, which
is real for this signature.  The 10 tensors (($3,1$) and ($1,3$)) belong to
the fundamental representation of $SO(5,5)$.  The fact that three-form field
strengths mix with their duals under the duality group is a higher-dimension analog of
electric-magnetic duality, and is a hallmark of an $S$ duality.  The fermions
belong to representations of the denominator algebra -- $SO(5) \times
SO(5)$ in this case -- in constructions of this type.

\subsection{Vacua with $(1,1)$ Supersymmetry}

Let us now consider vacua with non-chiral 2A supersymmetry.  In this case
the gravity supermultiplet consists of
\begin{equation}
\begin{array}{rl}
{\rm bosons:} &(3,3) + (3,1) + (1,3) + 4(2,2) + (1,1)\nonumber \\
{\rm fermions:} &2(3,2) + 2(2,3) + 2(2,1) + 2(1,2).
\end{array}
\end{equation}
In addition there can be massless vector supermultiplets, whose content is
\begin{equation}
(2,2) + 4(1,1) + 2(2,1) + 2(1,2).
\end{equation}
When there are $n$ Abelian vector multiplets, so that the gauge group is
$[U(1)]^n$, the duality group is $SO(4,n;\ZZ)$, and the moduli space ${\cal
M}_{4,n}$ is spanned by the $4n$ scalar fields belonging to the vector
multiplets.  The additional scalar field in the supergravity multiplet, which
can be identified as the dilaton, has a moduli space $\RR$, so that the
complete moduli space is ${\cal M}_{4,n} \times \RR$.

Toroidal compactification of the heterotic string to 6d gives a vacuum with $(1,1)$
supersymmetry and $n = 20$, which is the standard Narain result.\cite{narain86}
Note that the compactification of 20 left-moving dimensions gives rise to the
20 vector fields belonging to the vector supermultiplets, whereas the
compactification of 4 right-moving dimensions gives the 4 vector fields
belonging to the supergravity multiplet.

The heterotic string vacuum described above (with $n = 20$) 
has a dual description given by the type
IIA string compactified on $K3$.\cite{hull94}  
We won't present a complete discussion of
this result, which will be discussed by other lecturers, but simply verify
that the massless field content works out correctly.  Compactification on $K3$
breaks half the supersymmetry that is present in 10d, 
giving the desired $(1,1)$ supersymmetry, so it suffices to check
the bosonic field content.  To do this, we need some basic facts about the cohomology
and the moduli space of $K3$.  As usual for a compact connected 4-manifold,
$b_0 = 1$ and $b_4 = 1$. Furthermore $b_1=b_3=0$, so the
only non-trivial cohomology is $H^2$.  It has 22 generators, which can be
chosen to be 3 self-dual 2-forms $(b_2^+ = 3)$ and 19 anti-self-dual 2-forms
$(b_2^- = 19)$.  The moduli space of complex structure deformations, for a $K3$
of fixed volume, is ${\cal M}_{3,19}$, which is 57 dimensional.  
Including the volume of the manifold, there are 58 moduli
altogether.  This implies that the 6d zero modes obtained form the 10d metric consist of 
the 6d metric and 58 scalar fields.  Altogether, the massless bosonic IIA fields on $K3$ give
\begin{equation}
\begin{array}{rl}
g_{\mu\nu} & \rightarrow (3,3) + 58 (1,1)\nonumber \\
\phi & \rightarrow (1,1)\nonumber \\
A_\mu & \rightarrow (2,2) \nonumber \\
B_{\mu\nu} & \rightarrow (3,1) + (1,3) + 22(1,1)\nonumber \\
C_{\mu\nu\rho} & \rightarrow 23(2,2).
\end{array}
\end{equation}
One of the 23 vectors obtained from $C_{\mu\nu\rho}$ arises because a 3-form in
6d is equivalent (by a duality transformation) to a vector.  These
multiplicities agree precisely with those of $(1,1)$ supergravity coupled
to 20 vector supermultiplets.

The 6d duality between the heterotic string on $T^4$ and the IIA string on $K3$
can be lifted to a 7d duality.\cite{witten95a}  Since the IIA theory can be viewed as M
theory on $S^1$, it is plausible that one can identify this $S^1$ with one of the
$S^1$'s inside the $T^4$ used to compactify the heterotic theory.
Decompactifying this $S^1$ then leaves a 7d duality between M theory
compactified on $K3$ and the heterotic theory compactified on $T^3$.  This
duality also passes all the checks that have been made.  For example, the
duality group computed from both viewpoints is $SO(3,19; \ZZ)$ and the moduli
space is ${\cal M}_{3,19} \times \RR$.

It is even possible to go one more step and to lift this duality to 8d
using ``F theory.''\cite{vafa96a}  This topic is beyond the scope of these lectures, so let's
just state the result.  On the one hand, consider the heterotic string
compactified on $T^2$, which has a duality group $SO(2,18; \ZZ)$ and a moduli
space ${\cal M}_{2,18} \times \RR$.  The F theory dual description consists of a
non-perturbative type IIB vacuum, which can be formally described as a
compactification from 12d to 8d on a special class of $K3$
manifolds -- those that have an elliptic fibration. Remarkably, the moduli space of these
$K3$'s is given by a ${\cal M}_{2,18}$ subspace of the ${\cal M}_{3,19}$
moduli space of $K3$'s, so that the matching of moduli
spaces required for the duality again works beautifully.

The 6d duality between the heterotic string theory on $T^4$ and the IIA string
theory on $K3$ is an $S$ duality.  One way to see this is by comparing
low-energy effective supergravity Lagrangians and noting that the field mapping
between the two descriptions includes
\begin{equation}
\phi_H = - \phi_{IIA},
\end{equation}
as well as a duality transformation of the two-form potential.
This relation between dilaton fields implies that the coupling constants are
reciprocal to one another, which is $S$ duality.  This also means that the  
heterotic and type IIA strings, both of which occur in the 6d theory,
are electric--magnetic duals of one another.  When one is regarded as
fundamental,  the other must be viewed as a soliton.  This observation is
the key to understanding their 10d origins.  In the 10d IIA theory the magnetic
dual of the IIA string is the 5-brane.  Thus, the heterotic string in 6d
arises from wrapping a 5-brane of topology $K3 \times S^1$ on the
spatial $K3$, leaving a string $(S^1)$ in 6d.\cite{duff95c,harvey95a,townsend95d} 
The corresponding BPS soliton of the 6d supergravity field equations
has been constructed explicitly.\cite{sen95a,harvey95a}  
The converse story, which must surely be true too, is
less well established.  It requires that the IIA string in 6d should arise from
wrapping a 5-brane of the 10d heterotic theory on the spatial $T^4$.   

We have now found two entirely different constructions of heterotic strings as
M theory solitons.  The first one (discussed in Section 2.9)
arises from wrapping the M theory 2-brane
on $S^1/\ZZ_2 = I$.  The second one, which we have just found, entails wrapping
the M theory 5-brane on $K3$.  Later, we will discuss a class of vacua
in which both kinds of heterotic strings can occur at the same time.

In the heterotic picture, perturbative reasoning shows 
that there is enhanced gauge symmetry at
singular points of the moduli space ${\cal M}_{4,20}$.  It is
interesting to ask where the additional massless states come from in the dual
description.  The mechanism, which is non-perturbative,
is that the singular points of the moduli space correspond to
limits in which a two-cycle on the $K3$ shrinks to a point.\cite{witten95a,aspinwall95b}   
Type IIA 2-branes wrapped on the two-cycle give 0-branes in 6d whose mass is
proportional to the area of the two cycle.  There is an ADE classification of
the two-cycles on $K3$ that can vanish, which has just the properties required to
account for the symmetry enhancement that is 
obtained in the dual heterotic description.

We have explained that a type IIA vacuum in 6d should have a duality group $SO(4,n;\ZZ)$ and
a moduli space ${\cal M}_{4,n} \times \RR$, and then we presented a pair of dual
constructions for the special case $n = 20$.  It is natural to ask whether
other values of $n$ are also possible.  Constructions that give
other values of $n$ are conveniently described from the M theory viewpoint.
The $n = 20$ result, itself, can be viewed as arising from M theory compactified on
$K3 \times S^1$, since type IIA theory is M theory compactified on $S^1$.  
To generalize this, the idea is
to replace the compact space by $(K3 \times S^1)/\ZZ_h$.  For this to
work, one needs to restrict to a class of $K3$'s having a $\ZZ_h$ discrete
symmetry, and then to combine the action of the generator of this group with a
rotation by $2\pi/h$ on the circle.  This ensures that there are no fixed
points, so that a smooth manifold results.  The $h=2$ case was 
analyzed in ref.~\cite{schwarz95d}.
The possible discrete symmetries of
$K3$'s have been classified by the mathematician Nikulin.\cite{nikulin80}  
Chaudhuri and Lowe have applied his results to the problem at hand to
conclude that the complete set of possibilities is given by \cite{chaudhuri95a}
\begin{equation}
\begin{array}{rl}
h = 2 & \rightarrow n = 12\nonumber \\
h = 3 & \rightarrow n = 8\nonumber \\
h = 4 & \rightarrow n = 6\nonumber \\
h = 5,6 & \rightarrow n = 4\nonumber \\
h = 7,8 & \rightarrow n = 2\nonumber .
\end{array}
\end{equation}
Thus, there are consistent string vacua for these values of $n$.  The $h = 5$
and $6$ constructions give the same massless spectrum and moduli space, but it
is not known whether they are completely identical. (The same remark applies to
$h = 7,8.$)  It is also not known whether other $n$ values could be obtained by other
constructions that have not yet been considered.

In the special case of $h = 2$ a dual construction is known.\cite{chaudhuri95b}  One
starts with the $E_8 \times E_8$ heterotic string compactified on $T^4$, and
mods out by a $\ZZ_2$ that interchanges the two $E_8$'s.  Clearly,
this reduces the rank by 8 (reducing $n$ from 20 to 12).

\subsection{Vacua with $(2,0)$ Supersymmetry}

Type IIB supersymmetry in 6d admits two massless supermultiplets.  The gravity
supermultiplet particle content is
\begin{equation}
(3,3) + 4(2,3) + 5(1,3)
\end{equation}
and the tensor supermultiplet particle content is
\begin{equation}
(3,1) + 4(2,1) + 5(1,1).
\end{equation}
A superstring vacuum with IIB supersymmetry in 6d can be obtained by
compactifying the type IIB superstring on $K3$.  The resulting massless bosons
in 6d are as follows:
\begin{equation}
\begin{array}{rl}
g_{\mu\nu} & \rightarrow \ (3,3) + 58(1,1)\nonumber \\
\phi, \chi & \rightarrow \ 2(1,1)\nonumber \\
B_{\mu\nu}^{(1)}, B_{\mu\nu}^{(2)} & \rightarrow \ 2(3,1) + 2(1,3) +
44(1,1)\nonumber \\
A_{\mu\nu\rho\lambda}^+ & \rightarrow \ 3(1,3) + 19(3,1) + (1,1).
\end{array}
\end{equation}
This shows that the massless content is given by the gravity multiplet plus 21
tensor multiplets.

Since the particle content is chiral, there are potential gravitational
anomalies.  In six dimensions these are characterized by eight-forms.  Up to a
common overall normalization, the contributions of each of the chiral fields is
as follows:
\begin{equation}
\begin{array}{rl}
2(2,3): &{49\over 72} \tr R^4 - {43\over 288} (\tr R^2)^2\nonumber \\
\nonumber\\
2(1,2): &{1\over 360} \tr R^4 + {1\over 288} (\tr R^2)^2\nonumber \\
\nonumber\\
(1,3): &{7\over 90} \tr R^4 - {1\over 36} (\tr R^2)^2.
\end{array}
\end{equation}
Combining these with weights corresponding to one gravity multiplet and $n_T$
tensor multiplets, one finds that the sum vanishes provided that $n_T = 21$.  Thus,
$n_T = 21$, the result we found by $K3$ compactification of the IIB
superstring, is the only value that can give a consistent anomaly-free 
theory.\cite{townsend84b}

The vacua with $n_T = 21$ have the duality group $SO(5,21;\ZZ)$ and the moduli
space ${\cal M}_{5,21}$ parametrized by the 105 scalars in the tensor
multiplets.  Note that there are no scalars in the gravity supermultiplet, so
that this is the complete moduli space.  This also means that the dilaton must be
one of these 105 scalar fields.  Compactifying further on a circle
to 5d gives one more scalar field corresponding to the radius of the
circle.  There are no other new scalars, since the 6d theory has
no vector fields. Then the moduli space becomes ${\cal M}_{5,21} \times \RR$.  This 5d
model has the same supersymmetry, massless fields, and moduli space as the
heterotic string compactified to 5d, so it is natural to conjecture that they
are dual.  In fact, this is a consequence of two dualities that we have already
discussed:  Type IIA on $K3 \sim$ heterotic on $T^4$ and IIA on $S^1 \sim$ IIB
on $S^1$.  In the heterotic picture, the scalar field that corresponds to the
$\RR$ factor in the moduli space is the dilaton.  Thus, we have a $U$ duality:
the heterotic string compactified on $T^5$ at strong coupling corresponds to the type IIB
string on $K3 \times S^1$ at large radius of the $S^1$.  In particular, the
strong-coupling limit of the 5d heterotic string is six dimensional.  This is
analogous, of course, to the fact that the strong-coupling limit of the 10d type IIA string or
$E_8 \times E_8$ heterotic string is eleven dimensional.

The 6d theory with IIB supersymmetry and 21 tensor supermultiplets has another dual
description given by M theory compactified on the 
orbifold $T^5/\ZZ_2$.\cite{dasgupta95,witten95d}
The $\ZZ_2$ acts on each of the five circles of the torus, which introduces 32
orbifold points.  Including the other six space-time dimensions, they are 32
orbifold planes.  In units where an M theory 5-brane carries one
unit of magnetic charge, it turns out that each of these orbifold planes
carries $- 1/2$ unit of magnetic charge.  The charge cannot be cancelled
locally, but it can be cancelled globally by introducing 16 5-branes.  
This is necessary, since the total charge on a compact space must vanish. One can
then account for the massless field content as follows:  Compactification of
11d supergravity on $T^5/\ZZ_2$ gives an untwisted sector consisting of the
gravity multiplet and five tensor multiplets.  Each of the 5-branes
introduces an additional tensor multiplet, so that altogether there are 21 of them,
as required. The 5-branes can be represented as points on the $T^5/\ZZ_2$, and  their
coordinates are controlled by the five scalar fields in the tensor multiplet
associated to the 5-brane.  The 25 scalar fields belonging to the other five
tensor multiplets arise from zero modes of the 11d metric and three-form on
$T^5$.

The 6d theory contains, among other things, self-dual solitonic strings whose
tension can become arbitrarily small in suitable limits.  In terms of the IIB
superstring compactified on $K3$, their appearance can be traced to singular
limits in which a two-cycle on the $K3$ shrinks to a point.\cite{witten95e}
The reason for this is
that the self-dual 3-brane can wrap around the cycle, leaving a string in
the 6d space-time whose tension is proportional to the area of the two-cycle.
In the IIA case, the corresponding mechanism gave 0-branes with an ADE
classification.  Here it gives non-critical strings with an ADE classification.
The existence of strings whose tension can become small, so that they
effectively decouple from gravity, is an interesting phenomenon.  It would be
very desirable to have a better understanding of their properties, because they
seem to encode in a deep way the essential features of $N = 4$ gauge theory.  The point is
that, compactifying further to 4d on $T^2$,  windings of the string
around the two cycles of the torus give electric and magnetic charge in 4d.\cite{verlinde95}
The $SL(2,\ZZ)$ duality of $N=4$ gauge theory derives from that of the torus (again!).
The appearance of non-critical strings can also be understood in terms of the
description in terms of M theory compactified on $T^5/\ZZ_2$.\cite{witten95d}  
In this case, there are 16 ``parallel'' 5-branes, represented by points on the compact space.
Since an M theory 2-brane can be suspended between parallel
5-branes, when a pair of 5-branes approach one another this
2-brane is approximated by a string whose tension is proportional to the
separation of the 5-branes.

\section{6D $\!$STRING $\!$VACUA WITH ${\bf N=1}$ SUPERSYMMETRY}

The preceding section described various possibilities for superstring vacua
with extended supersymmetry in 6d.  The supersymmetry was very
constraining, so the classification presented was reasonably complete.  In the
case of $N = 1$ supersymmetry in 6d the story becomes much more
complex.  Each time the number of supersymmetries or the number of
uncompactified dimensions is decreased, new issues arise.  While our ultimate
goal is to understand vacua with $N = 1$ or $N=0$ in 4d, I am most
comfortable proceeding in steps, absorbing the lessons at one stage before
moving on to the next one.  The cutting edge, where the understanding is
increasing most rapidly at the present time, is for vacua with $N = 1$ in 6d
or with $N = 2$ supersymmetry in 4d.  I will only
discuss the former, and even this will not be complete.  Many, but not all, 4d
vacua with $N = 2$ can be obtained from these by a subsequent $T^2$
compactification.

$N = 1$ supersymmetry in 6d admits four kinds of massless supermultiplets:
\begin{equation}
\begin{array}{rl}
{\rm gravity}: & (3,3) + 2(2,3) + (1,3)\nonumber \\
{\rm tensor}: & (3,1) + 2(2,1) + (1,1) \nonumber \\
{\rm vector}: & (2,2) + 2(1,2) \nonumber \\
{\rm hyper}: & 2(2,1) + 4(1,1).
\end{array}
\end{equation}
In general, a 6d $N = 1$ string vacuum will give one gravity multiplet, $n_T$ tensor
multiplets, $n_V$ vector multiplets, and $n_H$ hyper multiplets.  Since all
models of this kind are chiral, anomaly cancellation always provides
non-trivial constraints.\cite{green84,green85} 
For example, cancellation of the $\tr R^4$ term in the
anomaly eight-form gives the requirement
\begin{equation}
n_H + 29n_T = n_V + 273.  \label{fivea}
\end{equation}

In this section we will discuss $N = 1$ models constructed in a number of
different ways.  One approach is compactification of the $SO(32)$ theory on a
smooth $K3$.  Such models can have non-perturbative symmetry enhancement when
5-branes are included.  A second approach is $K3$ compactification of the $E_8
\times E_8$ theory.  Non-perturbatively, this can be regarded as M theory
compactified on $K3 \times S^1 /\ZZ_2$.  In this case there is freedom associated
with dividing the instanton number between the two $E_8$'s as well as the
possibility of including 5-branes.  A third approach that we will mention,
which turns out to be dual to one of the $E_8 \times E_8$ compactifications, is
based on $T^4/\ZZ_2$ orbifold compactification of the $SO(32)$ theory.

\subsection{General Considerations}

For the most part we will consider models with $n_T = 1$, in which case eq.~(\ref{fivea})
simplifies to $n_H = n_V + 244$.  When $n_T = 1$ it is possible to give a
manifestly covariant effective action for the massless modes.  The point is
that the two-form with self-dual field strength in the gravity multiplet and
the two form with anti-self-dual field strength in the tensor multiplet can be
combined and represented by a two-form with an unconstrained field strength.
Another advantage of $n_T = 1$ is that anomaly cancellation can be achieved by
straightforward analogs of the techniques introduced for 10d models
in Ref.~\cite{green84}.  (Otherwise, a generalization given in Ref.~\cite{sagnotti92}
is required.)  In
the $n_T = 1$ case, with a semi-simple gauge group $G = \prod G_\alpha$, anomaly
cancellation is possible if the anomaly eight-form factorizes into a product of
two four-forms.  This means that $I_8 \sim X_{4} \wedge \tilde{X}_4$, where
\begin{equation}
X_4 = \tr R^2 - \sum_\alpha v_\alpha \tr F_\alpha^2
\end{equation}
\begin{equation}
\tilde{X}_4 = \tr R^2 - \sum_\alpha \tilde{v}_\alpha \tr F_\alpha^2.
\end{equation}
Here $F_\alpha$ is the Yang--Mills two-form associated to the group $G_\alpha$,
given by matrices in a convenient (fundamental, for instance) representation of
the Lie algebra.  The $v_\alpha, \tilde{v}_\alpha$ are numerical constants.

Anomaly cancellation is achieved by assigning non-trivial Yang--Mills and local
Lorentz gauge transformation assignments to the two-form $B_{\mu\nu}$, choosing
its field strength to be gauge invariant ($H = dB +$ Chern--Simons terms), and
adding a suitable counterterm of the form $\int B \wedge \tilde{X}_4$ to the
effective action.  Taking the exterior derivative of the field strength $H$
gives the Bianchi identity $dH \sim \tr R^2 - \sum v_\alpha \tr F_\alpha^2$.
Under the $S$-duality transformation
\begin{equation}
\phi \rightarrow - \phi, \quad H \rightarrow e^{-2\phi} * H,
\end{equation}
the Bianchi identity is intercharged with the one-loop corrected field
equations
\begin{equation}
d (e^{-2\phi} * H) \sim \tr R^2 - \sum \tilde{v}_\alpha \tr F_\alpha^2.
\end{equation}

When there are also $U(1)$ factors  in the gauge group, there can be additional
terms in the anomaly eight-form of the structure $F \wedge Y_6$, where $F$ is
the $U(1)$ field-strength two-form and $Y_6$ is a six-form.  When such terms
appear, anomaly cancellation can still be achieved provided there is a suitable
scalar field $\chi$ that transforms under the $U(1)$ gauge transformation
$(\chi \rightarrow \chi + \Lambda)$.  In this case its gauge-invariant field
strength has the structure $d\chi - A$.  This results in the $U(1)$ gauge field
eating the scalar $\chi$ to become massive.  Then there is no longer an
unbroken $U(1)$ gauge symmetry, but at least the theory is consistent. There is
an analogous mechanism in 4d, which has been known for a long time.\cite{witten84}
However, in 4d a scalar is dual to a 2-form, so that this is just a dual description of
the same mechanism as in 10d. In 6d that is not the case.

The constants $v_\alpha$ in the form-form $X_4$ have a simple interpretation,
pointed out in Ref.~\cite{duff96}, provided the group $\prod G_\alpha$ can be realized by a
perturbative heterotic string construction.  In this case the factor $G_\alpha$
is realized in the world-sheet theory as an affine Kac--Moody Lie algebra. For a
level $n_\alpha$ representation $v_\alpha$ is given by
\begin{equation}
v_\alpha \tr F_\alpha^2 = {n_\alpha\over h_\alpha} \Tr F_\alpha^2.
\end{equation}
We use the symbol `tr' for traces in the fundamental representation and `Tr' for
traces in the adjoint representation.
Here, $h_\alpha$ is the dual Coxeter number of the group $G_\alpha$.  In
practice, the only cases we will encounter are at level one $(n_\alpha = 1)$.
In this case one can show that $v = 2$ for an $SU(n)$ or $Sp(n)$ group, $v = 1$
for an $SO(n)$ group, $v = 1/3$ for $E_6$, $v = 1/6$ for $E_7$, and $v = 1/30$
for $E_8$.

We will mostly be interested here in $K3$ compactifications.  In this case,
integrating the four-form Bianchi identity over the $K3$ manifold gives a
consistency condition for the compactification.  Specifically, in the $SO(32)$
case, one obtains the condition
\begin{equation}
n_1 + n_5 = 24.
\end{equation}
Here, 24 arises is the Euler number of the $K3$ manifold and $n_1$ is the number
of instantons embedded in the $SO(32)$ gauge group, (\ie, the second Chern
class of the gauge bundle).  The integer $n_5$ is the number of 5-branes in
the solution.  These 5-branes correspond to delta-function sources in $dH$
at isolated points on the $K3$, filling the 6d space-time.  Their appearance is a
non-perturbative phenomenon.  In the case of $E_8 \times E_8$ models
compactified on $K3$, the integrated Bianchi identity gives a very similar
consistency condition
\begin{equation}
n_1 + n_2 + n_5 = 24.
\end{equation}
Here $n_1$ and $n_2$ denote the number of instantons embedded in each of the
two $E_8$ factors, and $n_5$ is again the number of (non-perturbative)
5-branes.  The study of branes in Section 2 showed that 5-branes of M theory
(or $E_8 \times E_8$ theory) and 5-branes of $SO(32)$ theory are quite
different.  This will be reflected here by the fact that inclusion of 5-branes
has very different implications for the 6d vacua in the two cases.

\subsection{K3 Compactification of the SO(32) Theory}

The $SO(32)$ theory can be viewed either as a heterotic string theory or a type
I string theory since the two descriptions are $S$ dual.  To make contact with
the interpretation of the constants $v_\alpha$ in the preceding subsection, the
heterotic interpretation is appropriate.  Let us begin with perturbative vacua
with $n_1 = 24$ and $n_5 = 0$, which were understood a long time ago.\cite{green85}

To describe the instantons in the $SO(32)$ gauge group, one
must select an $SU(2)$ subgroup in which to embed them.  One choice (but not the
only possible one, as we will see later) is to consider the decomposition
$SO(32) \supset SO (28) \times SU(2) \times SU(2)$ and embed the instantons in
one of the two $SU(2)$'s.  This leaves an unbroken $SO(28) \times SU(2)$ gauge
symmetry.  Using appropriate index theorems, one can compute the number of
hypermultiplet zero modes belonging to each representation of this group.  Such
an analysis gives the hypermultiplet content $10({\bf 28,2}) + 65({\bf 1,1})$.  Note that
altogether there are ${1\over 2} 28 \cdot27 + 3 = 381$ vector multiplets and $560 +
65=625$ hyper multiplets, which satisfies the condition $n_H = n_V + 244$.  The
contribution of each of these fields to the anomaly polynomial can be computed using
formulas in Refs.~\cite{green85,erler94,schwarz95c}.  One finds
\begin{equation}
X_4 = \tr R^2 - \tr F_1^2 - 2 \tr F_2^2
\end{equation}
\begin{equation}
\tilde{X}_4 = \tr R^2 + 2 \tr F_1^2 - 44 \tr F_2^2.
\end{equation}
Note that $v_1 = 1$ for the $SO(28)$ factor and $v_2 = 2$ for the $SU(2)$
factor, as expected for level-one representations.  The potential of scalar
fields has many flat directions.  At special values one can get symmetry
enhancement, the maximal case being $SO(28) \times [SU(2)]^6$.  However, the
generic situation is for symmetry breaking (``Higgsing'') to occur.  The gauge
symmetry can be broken to $SO(8)$ but not further.

Let us now go beyond perturbation theory and include 5-branes in the vacuum
configuration.  This problem was studied first by Witten, who identified the
5-branes as instantons that have shrunk to zero size, what
he called ``small instantons.''\cite{witten95c}  From the type I viewpoint the $SO(32)$
5-branes (as well as the heterotic strings) are $D$-branes.  In a type II
theory a single $D$-brane carries a $U(1)$ gauge symmetry, and when $n$ of them
coincide the group is enhanced to $U(n)$.  However, the projections that give a type I
theory modify these rules.  For example, the 32 coincident 9-branes that fill
the space-time are responsible for the $SO(32)$ gauge symmetry from the type I
viewpoint.  Dynamical 5-branes in the type I theory correspond to a group of
four stuck together from the type II viewpoint.  (This is the minimal unit, as
long as the compactification manifold is smooth.)  It turns out that such a
dynamical 5-brane carries a $Sp(1) = SU(2)$ gauge group, and that when $n$ of
them coincide the symmetry is enhanced to $Sp(n)$.\cite{witten95c}  
The number $n_5$ in the
condition $n_1 + n_5 = 24$ refers to the number of dynamical 5-branes, and
so the maximum number allowed is 24.

The example with the largest gauge group is achieved by taking $n_1 = 0$ and
$n_5 = 24$ and then taking the 24 5-branes to coincide.  This means that
they are all at the same point on the $K3$ manifold.  In this case the unbroken gauge
symmetry in 6d is $G = SO(32) \times Sp (24)$.  This group has rank
40, which is the world record for 6d models, as far as I know.  The massless
spectrum of this theory contains vector multiplets belonging to $SO(32) \times
Sp(24)$.  From the type I viewpoint, the $SO(32)$ vector multiplets arise as
zero modes of 99 open strings, \ie, open strings connecting 9-branes to
9-branes.  Similarly, the $Sp(24)$ vector multiplets arise from 55 open
strings, strings connecting 5-branes to 5-branes.  The massless hyper
multiplets turn out to be as follows:  55 strings give an $Sp(24)$
antisymmetric  tensor $({\bf 1,1127}) + ({\bf 1,1})$; 59 strings give 
${1\over 2} ({\bf 32,48})$; the $K3$
moduli give $20({\bf 1,1})$.  The factor of $1/2$ appears because it is possible to have
``half hyper multiplets'' when they belong to a pseudoreal representation of a
symmetry group.  (Fundamental representations of symplectic groups are
pseudoreal.)  Altogether there are 1672 vector multiplets and 1916
hypermultiplets, which again satisfies the condition $n_H = n_V + 244$.

The factorized anomaly polynomial of the $SO(32) \times Sp(24)$ 
model has~\cite{schwarz95c}
\begin{equation}
X_4 = \tr R^2 - \tr F_1^2
\end{equation}
\begin{equation}
\tilde{X}_4 = \tr R^2 + 2 \tr F_1^2 - 2 \tr F_2^2.
\end{equation}
The point to be noted is that the $SO(32)$ group has a perturbative heterotic
string interpretation, and it appears in $X_4$ with $v_1 = 1$, as expected.
The $Sp (24)$ group, on the other hand, is non-perturbative from the heterotic
string viewpoint, and it does not appear in $X_4$.

The ``small instanton'' model described above can be generalized in two ways.
One is to consider $n_5 < 24$ and to embed $n_1 = 24-n_5$ units of instanton
number in the $SO(32)$ group.  In this case the maximal non-perturbative gauge
group is $Sp(n_5)$.  The second generalization is to allow the $n_5$
5-branes to come apart into groups of $\{n_{5i}\}$ with  $\sum n_{5i} =
n_5$.  Then the non-perturbative gauge group is $\prod Sp(n_{5i})$.

We have discussed the significance of the $v_\alpha$'s in $X_4$, but not the
$\tilde{v}_\alpha$'s in $\tilde{X}_4$.  Sagnotti has shown that supersymmetry
considerations imply that the kinetic terms of the gauge fields have the form~\cite{sagnotti92}
\begin{equation}
\sum_\alpha (v_\alpha e^{-\phi} + \tilde{v}_\alpha e^\phi) \tr (F_\alpha \cdot
F_\alpha),
\end{equation}
where $\phi$ is the heterotic dilaton.  At weak coupling $(\phi \rightarrow -
\infty)$ the perturbative $v_\alpha$ term dominates.  However, as the coupling
is increased the second term becomes important.  In particular, if it happens
that $\tilde{v}_\alpha < 0$, then there is a singularity (divergent coupling
constant) at $\phi = \phi_0$, where
\begin{equation}
e^{2\phi_{0}} = - v_\alpha/\tilde{v}_\alpha.
\end{equation}
Duff and collaborators have argued that this singularity is associated with the
vanishing of a string tension.\cite{duff96b}  Specifically, they argue that there are
solitonic dyonic strings in 6d whose electric and magnetic charges $(p,q)$ are
proportional to $(v_\alpha, \tilde{v}_\alpha)$ and whose tension is given by
$T_{(p,q)} = pe^{-\phi} + qe^\phi$.  Thus, the tension of such a string
vanishes at the singularity.  In the examples that have been discussed so far
the $SO(n)$ group has $v_1 = 1$ and $\tilde{v}_1 = - 2$, and, therefore, its
coupling constant diverges for $e^{2\phi_{0}} = 1/2 = - p/q$.  Thus, a $(1,
-2)$ string becomes tensionless.  These theories are well-defined for weak
coupling $(e^{\phi_{0}} \ll 1)$ and should continue smoothly up to the
singularity at $e^{2\phi_{0}} = 1/2$.  At that point one expects a phase
transition to take place.  Beyond that point, our formulas are no longer
applicable.  Later, we will speculate about what happens at the phase
transition.

\subsection{K3 Compactification of the ${\bf E_8 \times E_8}$ Theory}

Perturbative vacua of the $E_8 \times E_8$ heterotic string compactified on $K3$
have $n_5 = 0$ (no 5-branes) and $n_1 + n_2 = 24$.  Let us begin by
considering the special case $n_1 = 24$, $n_2 = 0$, which corresponds to
embedding all 24 units of instanton number into one of the two $E_8$ factors.
A maximal subgroup of $E_8$ is $E_7 \times SU(2)$.  So if we embed the
instantons in this $SU(2)$, this leaves an unbroken $E_8 \times E_7$ gauge
symmetry.  In this case application of index theorems give massless
hypermultiplets transforming as $10({\bf 1,56}) + 65 ({\bf 1,1})$.  Just as in the $SO(28)
\times SU(2)$ model of the preceding subsection, there are 381 vector
multiplets and 625 hypermultiplets.  The factorized anomaly eight-form has
\begin{equation}
X_4 = \tr R^2 - {1\over 30} \tr F_1^2 - {1\over 6} \tr F_2^2
\end{equation}
\begin{equation}
\tilde{X}_4 = \tr R^2 + {1\over 5} \tr F_1^2 - \tr F_2^2.
\end{equation}
Note that $v_1 = 1/30$ and $v_2 = 1/6$ are the values expected for $E_8$ and
$E_7$, respectively.  By giving vevs to scalars corresponding to flat
directions one can find enhanced gauge symmetry as large as $E_8 \times E_7 \times
[SU(2)]^5$, or symmetry breaking giving a complete Higgsing of the $E_7$ leaving
only an unbroken $E_8$.  Note that the $E_8$ kinetic term becomes singular for
$\phi \rightarrow \phi_0$, where $e^{2\phi_{0}} = 1/6$.  At this point a
$(1,-6)$ string becomes tensionless.

The second possibility for perturbative vacua of the $E_8 \times E_8$ heterotic
string theory on $K3$ is to embed some instantons in each of the $E_8$'s.  It is
not possible to have $n = 1,2,3$, so the possibilities are
\begin{equation}
n_1 \geq n_2 = 24 - n_1 \geq 4.
\end{equation}
Embedding all instantons in $SU(2)$ subgroups can leave an unbroken $E_7 \times
E_7$.  The hypermultiplets in this case are ${1 \over 2}(n_1 - 4 )({\bf 56,1}) 
+ {1 \over 2} (n_2 - 4)({\bf 1,56}) + 62 ({\bf 1,1})$.  
Half integer coefficients are allowed, because
the {\bf 56} is a pseudoreal representation of $E_7$.  The factorized anomaly
polynomial is now
\begin{equation}
X_4 = \tr R^2 - {1\over 6} \tr F_1^2 - {1\over 6} \tr F_2^2
\end{equation}
\begin{equation}
\tilde{X}_4 = \tr R^2 + \left(1 - {n_1\over 12}\right) \tr F_1^2 + \left(1 -
{n_2\over 12}\right) \tr F_2^2. \label{fiveb}
\end{equation}

It is interesting to ask whether any of these models could be a dual
description of the perturbative $SO(32)$ string compactification.  Certainly
the groups we have found in the two cases are different.  However, in the case
of the $SO(32)$ model it was possible to Higgs to an $SO(8)$ subgroup  leaving
$X_4 = \tr R^2 - \tr F^2$ and $\tilde{X}_4 = \tr R^2 + 2 \tr F^2$, where $F$ refers
to the $SO(8)$.  In the case of the $E_7 \times E_7$ models under consideration
here, it is possible to completely Higgs one $E_7$ and to Higgs the other to
$SO(8)$.  Then, using the rule $\tr_{E_{7}} F^2 \rightarrow 6 \tr_{SO(8)} F^2$, we
are left with $X_4 = \tr R^2 - \tr F^2$ and $\tilde{X}_4 = \tr R^2 + \left( 6 -
{1\over 2}n_2\right) \tr F^2$.  Thus, the two $SO(8)$ models have the same
anomaly polynomials (and the same massless field content) for $n_1 = 16$, $n_2 =
8$.  Thus, it is plausible (and supported by other studies) that perturbative
$SO(32)$ compactifications and perturbative $(16,8)$ $E_8 \times E_8$
compactifications on $K3$ give the same moduli space of models.  Of course, the
portion of the moduli space that is visible in each approach is different.  To
find $SO(28)$ from the $E_8 \times E_8$ approach or $E_7 \times E_7$ from the
$SO(32)$ approach would require discovering the appropriate  unHiggsing.

Now, let us move beyond perturbation theory and consider $E_8 \times E_8$
models with 5-branes.  The 5-branes of $E_8 \times E_8$ are the
5-branes of M theory, which we saw carry a $ (2,0)$ tensor multiplet.
But the compactification on $K3$ cuts the supersymmetry in half leaving $N = 1$.
The $(2,0)$ tensor multiplet decomposes into a $N = 1$ tensor multiplet plus a
$N = 1$ hypermultiplet.  This is to be contrasted with the $SO(32)$
5-branes, which carry a $ (1,1)$ vector multiplet (that decomposes into a
$N = 1$ vector multiplet and a $N = 1$ hypermultiplet).  So $SO(32)$
5-branes carry vector multiplets, and that is why we found that they give
rise to additional (non-perturbative) gauge symmetry.  The $E_8 \times E_8$
5-branes, on the other hand, do not have vector multiplets and they do not
give additional gauge symmetry.  Rather, each $E_8 \times E_8$ 5-brane adds
a tensor multiplet (and a hypermultiplet).  Thus, by including them, we obtain
models with $n_T = n_5 + 1$ tensor multiplets.  When $SO(32)$ 5-branes
coincided, we found that the 55 strings connecting them gave massless gauge
bosons resulting in enhanced gauge symmetry.  When $E_8 \times E_8$ 5-branes
coincide, the 2-branes connecting them give tensionless strings.\cite{ganor96,seiberg96a}

Let's begin with the extreme case $n_5 = 24$, $n_1 = n_2 = 0$.\cite{seiberg96a}  
Since there are
no instantons to embed, the $E_8 \times E_8$ gauge symmetry is unbroken (so that $n_V
= 496$).  The number of tensor multiplets is $n_T = 25$.  The 24 5-branes
each give a hypermultiplet, and there are also 20 of them associated to the $K3$
moduli, so $n_H = 44$.  Note that these numbers satisfy
the anomaly condition $n_H + 29 n_T = n_V + 273$.

The $E_8 \times E_8$ theory -- viewed as M theory with two boundaries -- has no
anomalies in the bulk (where it is non-chiral), only on the boundaries.\cite{horava95}  The
anomaly cancellation condition, therefore, requires that the anomaly form be
expressible as a sum of two factorized pieces, one associated to each boundary.
This structure, which persists after $K3$ compactification, was analyzed by
Seiberg and Witten.\cite{seiberg96a}  
They found, in general, that for $n_1 + n_2 \leq 24$ and
$n_5 = 24 - n_1 - n_2$ the anomaly polynomial can be written in the form
\[
\left({1\over 2} \tr R^2 - A_1\right) \left({n_1 -8\over 4} \tr R^2 - {n_1 -
12\over 2} A_1\right)\]
\begin{equation}
+ \!\left({1\over 2} \tr R^2 - \!A_2\right) \!\!\left(\!{n_2 -8\over 4} \tr R^2 -
{n_2 -12\over 2} A_2\!\right)\!\!,
\end{equation}
where
\begin{equation}
A_i = \sum_\alpha v_{\alpha i} \tr F_{\alpha i}^2.
\end{equation}
Here $(n_1, A_1)$ and $(n_2, A_2)$ are associated to the two boundaries.
Remarkably, when $n_1 + n_2 = 24$, so that $n_T = 1$, this can be recast as a
single factorized expression.

One would like to have a global view of the moduli space of $N = 1$ vacua in
6d.  On the face of it, it would seem that there should (at least) be a
separate component for each possible number of tensor multiplets $n_T \leq 25$.
 The reason is that a tensor multiplet contains a tensor field $B_{\mu\nu}^-$
with an anti-self-dual field strength, and the only way such a field can
acquire mass is by joining up with another tensor field $B_{\mu\nu}^+$, whose
field strength is self-dual.  However, the only massless
$B_{\mu\nu}^+$ belongs to the gravity supermultiplet, and it must stay put if
the supersymmetry doesn't change.  This simple argument can be evaded, but this
requires something remarkable to happen.

In the M theory picture of the $E_8 \times E_8$ theory, we have argued that
extra tensor multiplets correspond to 5-branes in the bulk.  One could
imagine the number of 5-branes changing by emission or absorption by an
end-of-the-world 9-brane.\cite{seiberg96a}  Inside the 9-brane it can presumably turn into
an instanton.  To see the transition, one should consider a 5-brane in the
bulk very close to one of the 9-branes.  In this case a 2-brane which can
be suspended between them, is approximated by a string whose tension vanishes
as the 5-brane approaches the 9-brane.  So once again the proposed phase
transition is associated with the appearance of a string of vanishing tension.\cite{witten96a}
This is just what one needs to evade the argument in the preceding paragraph.
When a string goes to zero tension, all its modes go to zero mass, and this
undoubtedly includes an infinite number of massive $B_{\mu\nu}$ fields.  This
makes it possible for the $B_{\mu\nu}^-$ of the tensor multiplet to pair up
with a $B_{\mu\nu}^+$.  The idea is that $B_{\mu\nu}^{(n)}$ breaks up into
$B_{\mu\nu}^{(n)+}$ and $B_{\mu\nu}^{(n)-}$, but then, on the other side of the
transition, $B_{\mu\nu}^{(n)-}$ joins up with $B_{\mu\nu}^{(n+1)+}$ to become
massive again, leaving $B_{\mu\nu}^{(1)+}$ available to pair with
$B_{\mu\nu}^-$ from the tensor multiplet.  In view of the physical picture of
the transition in terms of 5-brane emmission and absorption from 9-branes
it seems likely that this actually happens and so there might be a single
connected moduli space of $N = 1$ vacua in 6d. Altogether, as required by eq.~(\ref{fivea}),
the massless tensor multiplet is replaced by 29 massless hypermultiplets.

\subsection{Models Without Phase Transitions}

We have seen that there are singularities, associated with the appearance of
tensionless strings, at specific value of the dilaton whenever one of the
$\tilde{v}_\alpha$ parameters is negative.  Indeed this phenomenon occurs in
almost all the models we have considered.  However, referring to eq.~(\ref{fiveb}), there
are no negative $\tilde{v}_\alpha$'s for the special case $n_1 = n_2 = 12$.
Thus the $(12,12)$ models, in which the instantons are embedded symmetrically
into the two $E_8$ factors could have smooth continuations from weak coupling
to strong coupling.\cite{duff96}  Specifically, this $E_7 \times E_7$ model lies on one
branch of an interesting moduli space of models.  These models, in general,
have a gauge group of the form $G = G_F \times G_D$, where $G_F$ is realized perturbatively
by ``fundamental'' heterotic strings and $G_D$ is realized non-perturbatively
by ``dual'' heterotic strings.  The specific example we have here is a somewhat
degenerate case, since it has
$G_F = E_7 \times E_7, ~{\rm and}~ G_D = 0$.

Where do these two kinds of heterotic strings come from?  Non-perturbatively,
the vacua we are considering correspond to M theory compactified on $K3 \times
S^1 /\ZZ_2$.  Recall that we found two different ways to make heterotic strings
in M theory: ~1) as a 2-brane suspended between end-of-the-world 9-branes,
or (equivalently) as a 2-brane wrapped on $S^1 /\ZZ_2$; ~2) as a 5-brane (of
topology $K3 \times S^1$) wrapped on $K3$.  The claim is that the first
construction gives the ``fundamental'' heterotic string with its associated
gauge group $G_F$, and the second one gives the ``dual'' heterotic string with
its associated gauge group $G_D$.

In this class of $(12,12)$ models there is an $S$ duality that interchanges the
role of the two strings.  This is reflected in the structure of the factorized
anomaly polynomial, which has
\begin{equation}
X_4 = \tr R^2 - \sum_\alpha v_\alpha \tr F_\alpha^2
\end{equation}
\begin{equation}
\tilde{X}_4 = \tr R^2 - \sum_i v_i \tr F_i^2,
\end{equation}
where $G_F = \prod G_\alpha$ and $G_D = \prod G_i$.  We will refer to these
models as DMW models, since they were introduced by Duff, Minasian, and Witten.\cite{duff96}
The parameters $v_\alpha$ and $v_i$ take the perturbative values (listed earlier)
for $G_\alpha$ and $G_i$, respectively.  Thus the $G_F$ field strengths do not
appear in $\tilde{X}_4$ and the $G_D$ field strengths do not appear in $X_4$.
The $S$ duality transformation $\phi \rightarrow - \phi, H \rightarrow e^{-2\phi}
* H$ interchanges the Bianchi identity and the field equation for  $H$.  This
means that it interchanges $X_4$ and $\tilde{X}_4$, and hence $G_F$ and $G_D$.
Thus in one picture $G_F$ is realized perturbatively and $G_D$ is realized
non-perturbatively, while after the $S$ duality transformation the situation is
reversed.  In this case the duality is called ``heterotic string -- heterotic
string duality.''\cite{aldazabal96}  
This duality is to be contrasted with the ``type IIA string
-- heterotic string duality'' discussed earlier.

By a remarkable coincidence, a dual type I construction of the same class of
models was discovered independently by Gimon and Polchinski,\cite{gimon96} 
and posted to the
hep-th archives on the same day as the Duff, Minasian, Witten paper.  
(For earlier related work see ref. \cite{pradisi}.) The GP
construction considers type I superstrings compactified on the orbifold
$T^4/\ZZ_2$.  This orbifold, which is a singular limit of a $K3$, has 16 fixed
points.  To make a consistent model, it is necessary to arrange for the
cancellation of certain tadpoles introduced by the orientifold projection used
to define the type I theory.  This requires the introduction of 32 Dirichlet
9-branes and 32 Dirichlet 5-branes.  The 9-branes would give an
$SO(32)$ gauge group in 10d, but after the compactification on $T^4/\ZZ_2$ to
6d it turns out that ``99 open strings'' can give at most a $U(16)$ gauge
group.  This can be Higgsed to various subgroups.  We will refer to the gauge
group arising in this way as $G_9$.  The 32 5-branes are required to clump
in groups of four (as in subsection 2), so they give eight dynamical
5-branes.

Recall the $SO(32)$ condition $n_1 + n_5 = 24$.  This is satisfied in the GP
model by $n_5 = 8$.  The reason that the model has $n_1 = 16$ is that each of the orbifold
points contains a ``hidden'' instanton, as can be demonstrated by blowing up
the singularity.\cite{berkooz96}  
However, this blow-up does not give the $n_1 = 16$, $n_5 = 8$
model described in subsection 2.  The reason for this is that the instantons
are embedded in the $SO(32)$ group differently than they were in the examples
discussed previously.  The relevant embedding uses the maximal subgroup $SO(4n)
\supset Sp (n) \times SU(2)$ for the case $n = 8$.  This accounts for the fact
that an $Sp(8)$ gauge group can be obtained when the 5-branes coincide away
from the orbifold points.  At an orbifold point it is possible to have one-half
of a dynamical 5-brane (which is two Dirichlet 5-branes).  This means that there
are $2^{15}$ topological sectors according to which of the orbifold points have
a half-integral number of 5-branes attached.  In fact, a $T$ duality
transformation (on the $T^4/\ZZ_2$) interchanges the 5-branes and
9-branes, so there are also $2^{15}$ topological sectors for the
9-branes, giving $2^{30}$ altogether.

It turns out that there are new non-perturbative anomalies that rule out most
of these topological sectors.  The issue, roughly, (see Ref.~\cite{berkooz96} for details)
is that these sectors would give states  belonging to the wrong spin(32)
conjugacy classes.  Consistency allows for there to be either 0, 8, or 16 half
5-branes attached to orbifold points and similarly for the 9-branes.
Thus, in view of the symmetry between them, there are altogether six
topologically distinct sectors.

Let us now consider the gauge groups that can be obtained in the GP
construction.  As we have seen before, if $n$ dynamical 5-branes coincide at
a non-singular point of $T^4/\ZZ_2$, the 55 open strings connecting them give a
$Sp(n)$ gauge group.  If, on the other hand, $m/2$ dynamical 5-branes
coincide at one of the $T^4/\ZZ_2$ orbifold points, the 55 open strings
connecting them turn out to give a $U(m)$ gauge group.  Thus, altogether, the
gauge group arising from 55 open strings is~\cite{gimon96}
\[
G_5 = \prod_I U(m_I) \cdot \prod_J Sp (n_J)\]
\begin{equation}
{1\over 2} \sum m_I + \sum n_J = 8.
\end{equation}
The largest group possible -- $U(16)$ -- is realized if all of the 5-branes
are at a single orbifold point.  This satisfies the non-perturbative criterion
given above.   The structure of $G_9$, the group given by 99 open strings is
exactly the same, as required by $T$ duality.  Thus altogether the gauge group of
a GP model is $G = G_5 \times G_9 \subseteq U(16) \times U(16)$.  The $U(1)$ factors
are broken by the mechanism described earlier,\cite{berkooz96} so actually
$G_{max} = SU(16) \times SU(16)$.

There are various massless hyper multiplets in the spectra of 99, 59, and 55
open strings.  Aside from $U(1)$ terms of the form $F \wedge Y_6$, they result
in an anomaly polynomial $X_{4} \wedge \tilde{X}_4$ with
\begin{equation}
X_4 = \tr R^2 - 2 \sum_\alpha \tr F_\alpha^2
\end{equation}
\begin{equation}
\tilde{X}_4 = \tr R^2 - 2 \sum_i \tr F_i^2,
\end{equation}
where, now, $G_5 = \prod G_\alpha$ and $G_9 = \prod G_i$.  The reason that the
coefficients are all $v_\alpha = v_i = 2$ is because the groups are unitary or
symplectic groups, for which $v = 2$ is the perturbative value.  Note that in
the GP construction the entire $G_5 \times G_9$ is realized perturbatively,
since both factors are associated with weakly coupled open strings.  Moreover,
the interchange of the two groups $G_5 \leftrightarrow G_9$ is achieved by a $T$
duality transformation.

Clearly the DMW and GP models are closely related, but what is the exact
correspondence?  In the DMW picture the two groups are carried by two kinds of
heterotic strings, related by an $S$ duality, whereas in the GP picture the two
groups are carried by two kinds of open strings, related by a $T$ duality.
Recall that in the 10d $SO(32)$ theory the group was also carried by either
heterotic or open strings.  In that case the heterotic string is a BPS soliton
of a type I theory, and the two descriptions are $S$ dual.  In the 6d problem being
considered now, the story is similar.  Both kinds of heterotic strings, F and
D, are solitons ($D$-branes, in fact) of the type I description.\cite{berkooz96}  One
corresponds to the 10d heterotic string and the other corresponds to the dual
10d 5-brane wrapped on the orbifold.  $T$ duality interchanges these two
solitons.  It is less straightforward to look for the open strings in the
heterotic construction since they are not BPS solitons.  Altogether, the DMW
and GP models are $U$ dual, since the mapping between them turns the $S$ duality of
the DMW picture into the $T$ duality of the GP picture.  This is an example of
``duality of dualities.''  These models also have dual descriptions in
terms of type IIA theory~\cite{aspinwall96} and F theory,\cite{morrison96} 
but since I've gone on long enough already, that will have to wait
for another occasion.

\end{document}